\documentclass[twocolumn,showpacs,preprintnumbers,amsmath,amssymb,superscriptaddress]{revtex4-1}   

\usepackage{graphicx}
\usepackage{dcolumn}
\usepackage{empheq}
\usepackage{amsmath,mathrsfs,amssymb,amsthm}  
\usepackage{hhline} 
\usepackage{wasysym,enumerate,color}
\usepackage{epstopdf} 
\usepackage{verbatim}  
\usepackage{hyperref}  
\usepackage{color}
\usepackage{xspace}
\usepackage{bm}
\usepackage{ulem} 

\usepackage{etoolbox}
\AtBeginEnvironment{array}{\setcounter{subeqn}{0}}
\newcounter{subeqn} %

\allowdisplaybreaks  

\renewcommand{\emph}[1]{\textit{#1}} 

\definecolor{darkgreen}{rgb}{0,0.45,0}
\definecolor{purple}{rgb}{0.6,0,0.5}
\definecolor{orange}{rgb}{1,0.5,0}
\definecolor{darkred}{rgb}{.7,0,0}
\definecolor{darkblue}{rgb}{0,0,.3}
\definecolor{grey}{rgb}{.6,.6,.6}
\definecolor{dimgreen}{rgb}{0.2,0.6,0.1}
\hypersetup{colorlinks,linkcolor=blue,urlcolor=blue,citecolor=blue}

\newcommand{\backup}[1]{}
\def\Nkept{\ensuremath{N^\ast_\mathrm{kept}}\xspace}

\newcommand{\nd}{n_d}
\newcommand{\ndsigma}{n_{d\sigma}}

\newcommand{\md}{m_d}
\newcommand{\HFL}{H_{\rm FL}}
\newcommand{\imp}{{\rm imp}}

\newcommand{\Tk}{T_K}

\newcommand{\Tkaf}{T_K^{\rm (af)}}
\newcommand{\Tkchi}{T_K}
\newcommand{\CA}{C_{\! A}}
\newcommand{\tCA}{\widetilde C_{\!A}}
\newcommand{\CV}{C_{V}}
\newcommand{\CT}{C_{T}}
\newcommand{\cA}{c_{A}}
\newcommand{\tcA}{\tilde c_{A}}
\newcommand{\tBA}{\widetilde B_A}
\newcommand{\cV}{c_{V}}
\newcommand{\cT}{c_{T}}
\newcommand{\Estar}{E_\ast}
\newcommand{\nzero}{n^0}
\newcommand{\tG}{\widetilde G}

\newcommand{\ezero}{{\varepsilon_0}} 
\newcommand{\ve}{\varepsilon} 
\newcommand{\Bzero}{{B_0}}

\newcommand{\muzero}{{\mu_0}} 
\newcommand{\muzerosigma}{{\mu_{0\sigma}}} 
\newcommand{\muzerosigmaprime}{{\mu_{0\sigma'}}} 
\newcommand{\ezerosigma}{{\varepsilon_{0\sigma}}} 
\newcommand{\dzerosigma}{{\delta_{0\sigma}}} 
\newcommand{\edsigma}{{\varepsilon_{d\sigma}}} 
 
\newcommand{\ed}{{\varepsilon_d}} 
\newcommand{\ted}{{\tilde \varepsilon_d}}

\newcommand{\balpha}{\overline \alpha}
\newcommand{\bphi}{\overline \phi}

\newcommand{\tes}{\tilde \varepsilon_{d\sigma}} 

\newcommand{\trhos}{\tilde \rho_\sigma}

\newcommand{\dels}{\delta_{0\sigma}}
\newcommand{\tds}{\tilde \Delta_\sigma}

\newcommand{\tzs}{\tilde z_\sigma}

\newcommand{\tRs}{\tilde R_\sigma}
\newcommand{\tIs}{\tilde I_\sigma}
\newcommand{\tRso}{\tilde R_{\omega\sigma}}
\newcommand{\tIso}{\tilde I_{\omega\sigma}}
\newcommand{\tRsV}{\tilde R_{V\! \sigma}}

\newcommand{\CAs}{C_{\! A \sigma}}
\newcommand{\tCAs}{\widetilde C_{\!A \sigma}}

\newcommand{\Eq}[1]{Eq.~(\ref{#1})} 

\newcommand{\Eqs}[1]{Eqs.~(\ref{#1})}

\usepackage{color} 
\newcommand{\be}{\begin{equation}}
\newcommand{\ee}{\end{equation}}

\begin{document}
\title{At which magnetic field, exactly, does the Kondo resonance begin to split? \\ A Fermi liquid description of the low-energy properties of the Anderson model}
\author{Michele Filippone}
\affiliation{Dahlem Center for Complex Quantum Systems and Institut f\"ur Theoretische Physik,
Freie Universit\"at Berlin, Arnimallee 14, 14195 Berlin, Germany}
\affiliation{Department of Quantum Matter Physics, University of Geneva, 24 Quai Ernest-Ansermet, CH-1211 Geneva, Switzerland} 
\author{C\u at\u alin Pa\c scu Moca}
\affiliation{BME-MTA Exotic Quantum Phase Group, Institute of Physics, Budapest University of Technology and Economics,
H-1521 Budapest, Hungary}
\affiliation{Department of Physics, University of Oradea, 410087, Oradea, Romania}
\author{Andreas Weichselbaum}
\affiliation{CM PMS Division, Brookhaven National Laboratory, Upton, New York 11973, USA}
\affiliation{Physics Department, Arnold Sommerfeld Center for Theoretical Physics and Center for NanoScience, Ludwig-Maximilians-Universit\"at M\"unchen, 80333 M\"unchen, Germany}
\author{Jan von Delft}
\affiliation{Physics Department, Arnold Sommerfeld Center for Theoretical Physics and Center for NanoScience, Ludwig-Maximilians-Universit\"at M\"unchen, 80333 M\"unchen, Germany}
\author{Christophe Mora}
\affiliation{Laboratoire Pierre Aigrain, \'Ecole normale sup\'erieure, PSL Research University, CNRS, Universit\'e Pierre et Marie Curie, Sorbonne Universit\'es, Universit\'e Paris Diderot, Sorbonne Paris-Cité, 24 rue Lhomond, 75231 Paris Cedex 05, France}
\date{\today}

\begin{abstract}
This paper is a corrected version of Phys.\ Rev.\ B
    \textbf{95}, 165404 (2017), which we have retracted because it
    contained a trivial but fatal sign error that lead to incorrect
    conclusions. --- We extend a recently-developed Fermi-liquid (FL)
  theory for the asymmetric single-impurity Anderson model [C. Mora
  \textit{et al.}, Phys.~Rev.~B, \textbf{92}, 075120 (2015)] to the
  case of an arbitrary local magnetic field.  To describe the system's
  low-lying quasiparticle excitations for arbitrary values of the bare
  Hamiltonian's model parameters, we construct an effective low-energy
  FL Hamiltonian whose FL parameters are expressed in terms of the
  local level's spin-dependent ground-state occupations and their
  derivatives with respect to level energy and local magnetic
  field. These quantities are calculable with excellent accuracy from
  the Bethe Ansatz solution of the Anderson model.  Applying this
  effective model to a quantum dot in a nonequilibrium setting, we
  obtain exact results for the curvature of the spectral function,
  $c_A$, describing its leading $\sim \varepsilon^2$ term, and the
  transport coefficients $c_V$ and $c_T$, describing the leading
  $\sim V^2$ and $\sim T^2$ terms in the nonlinear differential
  conductance. A sign change in $c_A$ or $c_V$ is indicative of a
  change from a local maximum to a local minimum in the spectral
  function or nonlinear conductance, respectively, as is expected to
  occur when an increasing magnetic field causes the Kondo resonance
  to split into two subpeaks. We find that
    the fields $B_A$, $B_T$ and $B_V$ at which $c_A$, $c_T$ and $c_V$
    change sign, respectively, are all of order $\Tk$, as expected,
    with $B_A = B_T = B_V = 0.75073 \Tk$ in the Kondo limit. 
\end{abstract}


\maketitle

\section{Introduction}    
 
The Kondo effect, arising from the exchange interaction between a
localized spin and delocalized conduction band, is characterized by a
crossover between a fully-screened singlet ground state and a free
local spin at energies well above the Kondo temperature scale
$T_K$.
  One of the most striking signatures of the Kondo effect is the
  occurrence of a sharp resonance near zero energy in the
  zero-temperature local spectral function $A(\ve)$, which splits
  apart into two subresonances when a local magnetic field $B$ is
  applied. Consequences of this Kondo peak and its field-induced
  splitting have been directly observed in numerous experimental
  studies of quantum dots tuned into the Kondo regime, where it causes
  a zero-bias peak in the nonlinear differential conductance $G(V)$,
  which splits into two subpeaks with increasing field.  Indeed, the
  observation of a field-split zero-bias peak has come to be regarded
  as one of the hallmarks of the Kondo effect in the context of
  transport through quantum dots~\cite{goldhaber1998,kogan2004,
    houck2005,amasha2005,quay2007,jespersen2006,liu2009}.  
  
  A minimal model for describing such experiments
  ~\cite{glazman1988,ng1988,glazman2003,*glazman2004,chang2009} is the
  two-lead, nonequibrium, single-impurity Anderson model, describing a
  ``dot'' level with local interactions that hybridizes with two leads
  at different chemical potentials.  Within the framework of this
  model (and its Kondo limit), numerous numerical and approximate
  analytical studies have explored the field-induced splitting of the
  Kondo peak in $A(\ve)$ and of the zero-bias peak in $G(V)$
  \cite{rosch2003,kehrein2005,doyon2006,Silva2008a,schoeller2009,Eckel2010,moca2011,gezzi2007,metzner2012,Pletyukhov2012,smirnov2013,schiller1998,boulat2008,Anders2005,anders2008,Cohen2014a,Dorda2015}.
  However, no \textit{exact}, quantitative description exists for how
  these splittings come about \footnote{An attempt in this direction
    was made in Ref.~\cite{hewson2005} for the symmetric Anderson
    model, using renormalized perturbation theory.  However their
    results disagree with the results presented here (see the end of
    Section~\ref{footnote:hewson2005}).}.  For example, it is natural
  to expect that the emergence of split peaks is accompanied by a
  change of the curvatures $\partial_\ve^2 A(\ve)|_{\ve=0}$ and
  $\partial_V^2 G(V)|_{V=0}$ from negative to positive
  \cite{hewson2005}.  A quantitative theory should yield exact results
  for the values of the ``splitting fields'', say $B_A$ and $B_V$,
  respectively, at which these curvatures change sign. This
  information would be useful, for example, as benchmarks against
  which future numerical work on the nonequilibrium Anderson model
  could be tested.

In the present paper, we use Fermi-liquid (FL) theory to compute these
quantities exactly within the context of the two-lead, single-impurity
Anderson model, for arbitrary particle-hole asymmetry. 
We develop an exact FL description of the low-energy
regime where both the temperature $T$ and the source-drain voltage $V$
are much smaller than a crossover scale $\Estar$, while the magnetic
field $B$ and the local level energy $\ed$ can be arbitrary. (In the
local-moment regime at zero field, $\Estar$ corresponds to the Kondo
temperature $\Tk$.)  Though our theory does not capture the full shape
of $A(\ve)$ for arbitrary $\ve$ or $G(V)$ for arbitrary $V$, 
it does describe their curvatures at zero energy and voltage, 
respectively, for \textit{arbitrary} values of $B$ and $\ed$.

FL theories for quantum impurity systems have been originally
  introduced by
  Nozi\`eres~\cite{nozieres1974,*nozieres1974b,*nozieres1978} with
  phenomenological quasiparticles, and by Yamada and Yosida on a
  diagrammatic
  basis~\cite{yosida1970,*yamada1975a,*yosida1975,*yamada1975b}. Later
  these theories were extended to orbital degenerate Anderson
  models~\cite{yoshimori1976,mihaly1978,sakano2011,sakano2011b,horig2014,Hanl2014},
  or extended in a renormalized perturbation
  theory~\cite{hewson1997kondo,hewson2005b,hewson1993,*hewson1993b,*hewson1994,*hewson2001,*hewson2004,*hewson2006b,*bauer2007,*edwards2011,*edwards2013,*Pandis2015,hewson2005,hamad2015,hewson2006}.
  They have also been extended to higher order terms in the low energy
  perturbative
  expansion~\cite{lesage1999a,*lesage1999b,freton2014}. The FL
  approach used here to obtain the above results builds on a recent
  formulation by some of the present authors of a Fermi-liquid theory
  for the single-impurity Anderson model \cite{mora2015}, similar in
  spirit to the celebrated FL theory of Nozi\`eres for the Kondo model
  \cite{nozieres1974,*nozieres1974b,*nozieres1978}.  One useful
  feature of FL approaches
  ~\cite{mora2015,nozieres1974,*nozieres1974b,*nozieres1978,oguri2001,glazman2005,sela2006,golub2006,gogolin2006,mora2008,mora2009a,mora2009,vitushinsky2008,fujii2010}
  is that they provide exact results for the nonlinear conductance in
  out-of-equilibrium settings, albeit only in the limit that
  temperature and voltage are small compared to a characteristic FL
  energy scale $\Estar$.  For example, in Ref.~\cite{mora2015} we
  obtained exact results for the differential conductance and the
  noise of the Anderson model for arbitrary particle-hole
  asymmetry~\cite{mora2015}, but zero magnetic field.  The FL
  parameters of this effective theory were written in terms of ground
  state properties which are computable semi-analytically using Bethe
  Ansatz, or numerically via numerical renormalization group (NRG)
  calculations \cite{Wilson1975,Bulla2008,Weichselbaum2012a}. We here
  extend this FL approach to arbitrary magnetic fields. This enables
  us to obtain exact results for the low-energy behavior of the
  spectral function and the nonlinear conductance for any $B$ and
  $\ed$, and to explore the crossovers from the strong-coupling
  (screened-singlet) fixed point to the weak-coupling (free-spin)
  fixed point of the Anderson model as functions of both these
  parameters.

Our work is based on the fact that the Kondo ground state remains a
Fermi liquid at finite magnetic field, as has been demonstrated by NRG
in Ref.~\onlinecite{Garst2005}.  There the Korringa-Shiba relation on the
spin susceptibility was shown to hold at arbitrary field, indicating
that the low-energy excitations above the ground state are
particle-hole pairs, as predicted by FL theory. Indeed, for 
both the Kondo model and the Anderson model, there is a
fundamental difference between a non-vanishing local magnetic field
and other perturbations such as temperature or voltage.
  Electrons conserve their spin after
scattering and are thus not sensitive to the chemical potential of the
opposite spin species. At zero temperature and bias voltage there is
no room for inelastic processes, regardless of the value of the
magnetic field, hence scattering remains purely elastic even when the
Kondo singlet is destroyed due to the applied field. In contrast, 
increasing temperature or voltage open inelastic channels by
deforming the Fermi surfaces of itinerant electrons. 

The rest of the paper is organized as follows. Sec.~\ref{sec-FL}
develops our FL theory for the asymmetric Anderson model
at arbitrary local magnetic field and shows how the FL parameters can be
expressed in terms of local spin and charge susceptibilities.  In
Sec.~\ref{sec:spectral-function-conductance}, we exploit the effective
FL Hamiltonian to
compute  two FL coefficients. $\tcA$ and $\cA$,
characterizing the zero-energy height and curvature of the spectral
function, respectively, and two FL transport coefficients, $\cT$ and
$\cV$, characterizing the curvatures of the conductance as function of
temperature $T$ and bias voltage $V$. From these 
we extract the splitting fields
$B_A$, $B_T$ and $B_V$ at which $c_A$, $c_T$ and $c_V$ change
sign, respectively.
This is done first at particle-hole symmetry,
then for general particle-hole asymmetry. 
 Sec.~\ref{sec-conclusion} presents a summary and conlusions. 

Our main physical result is that 
  throughout the local-moment regime the three splitting fields are
  all of order of the Kondo temperature $\Tk$, as expected. In the
  Kondo limit in which the Anderson model maps onto the Kondo model,
  we find $B_A = B_T = B_V = 0.75073 \Tk$, where $\Tk$ is defined from
  the zero-field spin susceptibility. 

  In a previous version of this paper, \cite{Filippone2017}, we had
  erroneously reached a different conclusion for $B_V$. We had
  summarized our main physical conclusions as follows in the abstract:
  ``Surprisingly, we find that the fields $B_A$ and $B_V$ at which
  $c_A$ and $c_V$ change sign are parametrically different, with $B_A$
  of order $T_K$ but $B_V$ much larger. In fact, in the Kondo limit
  $c_V$ never vanishes, implying that the conductance retains a (very
  weak) zero-bias maximum even for strong magnetic field and that the
  two pronounced finite-bias conductance side peaks caused by the
  Zeeman splitting of the local level do not emerge from zero bias
  voltage.''  These conclusion are simply incorrect --- they arose
  from a trivial but fatal sign error in Eq.~(25) during the
  computation of the conductance. When correcting this sign error and
  all its consequences, one finds that $B_V$, just as $B_A$ and $B_T$,
  is of order $\Tk$ throughout the local moment regime, as stated
  above.  We have therefore retracted Ref.~\cite{Filippone2017}; the
  present paper constitutes a corrected version thereof. We would like
  to emphasize, though, that the FL theory presented in section II of
  Ref.~\cite{Filippone2017} remains unchanged --- the sign error arose
  only during the \textit{application} of our FL to the computation of
  the conductance in section III. Indeed, the validity of our FL
  theory has been confirmed in a very recent series of three papers by
  Oguri and Hewson \cite{Oguri2017,*Oguri2017a,*Oguri2017b}, who
  presented a microscopic derivation of our FL relations using Ward
  identities. Their analysis pinpointed a likely source of error in
  our computation of the conductance in Ref.~\cite{Filippone2017},
  which indeed lead us to discover our sign error. Our corrected
  results for $c_T$ and $c_V$ fully agree with theirs (see
  appendix~\ref{app:oh}).

In Ref.~\cite{Filippone2017}, we had made an attempt to back up
  our FL predictions by using NRG to compute the equilibrium spectral
  function $A(\omega)$, in order to extract $c_A$ and $\tilde c_A$
  from its the low-frequency behavior.  
  In retrospect, that analysis had been unreliable --- in the
 regime of current interest, where $c_A$ and $\tilde c_A$ change sign and
 hence are very small, it is very challenging to
extract them accurately, and we had not exercised sufficient care
in doing so. We have now made another attempt, 
using a different NRG code that exploits all available symmetries,
allowing us to increase the number of
states kept during NRG truncation by a factor 7,
leading to significantly more accurate numerical results.
Moreover, we have  modified our strategy
for determining curvature coefficients from
discrete spectral data to directly extract $C_T$ and $C_V$ 
(rather than $c_A$ and $\tilde c_A$). Our new NRG results,
presented in Appendix~\ref{app:NRG}, 
are consistent with our corrected FL predictions
for $C_T$, $C_V$, $B_T$ and $B_V$.

\section{Fermi-liquid theory}\label{sec-FL}

\subsection{Anderson model}

The single-impurity Anderson model is a prototype model for magnetic
impurities in bulk metals or for quantum dot nanodevices, and more
generally for studying strong correlations in those systems. It
describes an interacting spinful single level tunnel-coupled to a
Fermi sea of itinerant electrons. Its Hamiltonian takes the form
\begin{equation}\label{am}
\begin{split}
H &= \sum_{\sigma,k} \varepsilon_{k} 
 c^\dagger_{k\sigma} c_{k\sigma}
+  \sum_\sigma \varepsilon_{d\sigma} 
\, \hat{n}_{d\sigma} \\
& + U \hat{n}_{d\uparrow} \hat{n}_{d\downarrow}
+ t \sum_{k,\sigma} \left( c_{k\sigma}^\dagger  d_\sigma +  d_\sigma^\dagger
c_{k\sigma} \right).
\end{split}
\end{equation}
Here $d^\dagger_\sigma$ creates an electron with spin $\sigma$ in
  a localized level with occupation number
  $\hat n_{d\sigma} = d_\sigma^\dagger d_\sigma$, spin-dependent
  energy $\varepsilon_{d\sigma} = \varepsilon_d - \sigma B/2$, local
  magnetic field $B$, and Coulomb penalty $U$ for double occupancy.
  $c^\dagger_{k \sigma}$ creates an electron with spin $\sigma$ and
  energy $\varepsilon_k$ in a conduction band with linear spectrum and
  constant density of states $\nu_0$ per spin species.  The local
  level and conduction band hybridize, yielding an escape rate
  $2 \Delta = 2 \pi \nu_0 t^2$.

  We will denote the ground state chemical potential for electrons of spin
  $\sigma$  by $\muzerosigma$. Although $\muzero_\uparrow $
  and $\muzero_\downarrow$ are usually taken equal, they formally are
  independent parameters that can be chosen to differ, 
  because the model contains no spin-flip terms, hence spin-up and
  -down chemical potentials have no way to equilibrate. In this paper,
  we will consider only the limit of infinite bandwidth~\footnote{In the non-universal case of a finite
      bandwidth, the Fermi liquid relations that we derive are no longer strictly valid. Nevertheless,
     the corrections to our predictions are expected to be
      small with the ratio of the maximum of $\Delta$, $B$, $|\varepsilon_d|$
      and $U$ over the bandwidth of the model}. Then
  $\muzero_\uparrow$ and $\muzero_\downarrow$ constitute the only
  meaningful points of reference for the model's single-particle
  energy levels.  Thus, ground state properties can depend on
  $\edsigma$ and $\muzerosigma$ only in the combination
  $\edsigma - \muzerosigma$, implying that they are invariant under
  shifts of the form
\begin{eqnarray}
\label{eq:general-shifts}
\edsigma \to \edsigma + \delta \mu_\sigma \, , 
\qquad \muzerosigma \to \muzerosigma + \delta \mu_\sigma \, . 
\end{eqnarray}
In Ref.~\onlinecite{mora2015}, this invariance was exploited for
spin-independent shifts ($\delta \mu_\sigma = \delta \mu$) when
devising a FL theory around the point $B=0$. Here we will exploit the
fact that the invariance holds also for spin-dependent shifts
to generalize the FL theory to arbitrary $B$.
Having made this point, we henceforth 
take $\muzero_\uparrow = \muzero_\downarrow = 0$ (but for clarity
nevertheless display $\muzerosigma$ explicitly in some formulas).  The
model's zero-temperature, equilibrium properties are then fully
characterized by $U$, $\Delta$, $\varepsilon_d$ and $B$.

\subsection{General strategy of FL theory \`a la Nozi\`eres}

Despite exhibiting strong correlations by itself, the ground state of
the Anderson model~\eqref{am} is a Fermi li\-quid for all values of $U$,
$\varepsilon_d$, $\Delta$ and $B$. A corresponding FL theory \`a la
Nozi\`eres was developed in \cite{mora2015} for small fields.  We now
briefly outline the general strategy used there, suitably adapted to
accommodate arbitrary values of $B$. Details follow in subsequent
sections.

The low-energy behavior of a quantum impurity model with a FL ground
state can be understood in terms of weakly interacting quasiparticles,
characterized by their energy $\varepsilon$, spin $\sigma$,
distribution function $n_\sigma (\varepsilon) $, and the phase shift
$\delta_{\sigma}(\varepsilon, n_{\sigma'})$ experienced upon
scattering off the screened impurity. At zero temperature, the
quasiparticle distribution reduces to a step function,
$n^0_{\muzerosigma}(\varepsilon) = \theta(\muzerosigma-\varepsilon)$,
and the phase shift at the chemical potential, denoted by
$\delta_{0\sigma} =
\delta_{\sigma}(\muzerosigma,n^0_{\muzerosigmaprime})$,
is a characteristic property of the ground state. It is related to the
impurity occupation function,
$n_{d\sigma} = \langle \hat{n}_{d\sigma} \rangle$, via Friedel's sum
rule, $\delta_{0\sigma} = \pi n_{d \sigma}$. Likewise, derivates of
$\delta_{0\sigma}$ w.r.t.\ $\varepsilon_d$ and $B$ are related to the
ground state values of the local charge and spin susceptibilities. The
ground-state dependence of local observables such as $n_{d\sigma}$ and their
derivatives on the model's bare parameters $U$, $\Delta$,
$\varepsilon_d$ and $B$ is assumed to be known, e.g.\ from Bethe
Ansatz or numerics.

The goal of a FL theory is to use such ground state information to predict
the system's behavior at non-zero but low excitation energies.
The weak residual interactions between low-energy quasiparticles can
be treated perturbatively using a phenomenological effective
Hamiltonian, $\HFL$, whose form is fixed by general symmetry
arguments. The coupling constants in $\HFL$, together with
$\delta_{0\sigma}$, are the ``FL parameters'' of the theory. The
challenge is to express these in terms of ground state properties,
while ensuring that the theory remains invariant under the shifts of
Eq.~\eqref{eq:general-shifts}. To this end, $\HFL$ is constructed in a
way that is \textit{independent} of $\muzerosigma$: it is expressed in
terms of excitations relative to a \textit{reference} ground state
with distribution $\nzero_\ezerosigma$ and spin-dependent chemical
potentials $\varepsilon_{0\sigma}$ chosen at some arbitrary values
close to but not necessarily equal to $\muzerosigma$.  The FL
parameters are then functions of $U$, $\Delta$ and the energy
differences $\edsigma - \ezerosigma$.  Importantly, and in keeping
with their status of depending only on ground state properties, they
do not depend on the actual quasiparticle distribution functions
$n_\sigma$, which are the only entities in the FL theory that depend
on the actual chemical potential and temperature. 

$\HFL$ is used to calculate $\delta_\sigma(\varepsilon,n_{\sigma'})$
for a general quasiparticle distribution $n_{\sigma'}$, to lowest
non-trivial order in the interactions. The result amounts to an
expansion of the phase shift in powers of
$\varepsilon -\varepsilon_{0\sigma}$ and
$\delta n_{\sigma} = n_\sigma - n^0_{\varepsilon_{0\sigma}}$, which
are assumed small.  Since the reference energies
$\varepsilon_{0\sigma}$ are dummy variables on which no physical
observables should depend, this expansion must be \textit{independent}
of $\varepsilon_{0\sigma}$.  This requirement leads to a set of
so-called ``Fermi liquid relations'' between the FL parameters, which
can be used to express them all in terms of various local ground state
observables, thereby completing the specification of $\HFL$.  Finally,
$\HFL$ is used to calculate transport properties at low temperature
and voltage.
 
\subsection{Low-energy effective model}

The phenomenological FL Hamiltonian has the form 
\begin{subequations}
\label{eq:H_FL}
 \begin{eqnarray}
\HFL & =&  \sum_\sigma \int_\varepsilon \, \varepsilon \, b_{\varepsilon
  \sigma}^\dagger b_{\varepsilon \sigma} +H_\alpha + H_\phi +\dots \, 
\label{eq:H_FL-expansion}
\\
\label{eq:H_alpha}
H_\alpha&=& - \! \sum_\sigma \int_{\varepsilon_{1},\varepsilon_2} 
\!\!
\bigg [ \frac{\alpha_{1\sigma}}{2\pi}  \bigl({\varepsilon_1 + \varepsilon_2 -2 \varepsilon_{0\sigma}}\bigr) \\
&&  \qquad + \frac{\alpha_{2\sigma}}{4\pi}\bigl({\varepsilon_1 + \varepsilon_2  -2 \varepsilon_{0\sigma}}\bigr)^2 \bigg]  
 b_{\varepsilon_1 \sigma}^\dagger b_{\varepsilon_2 \sigma} \, , 
\nonumber \\
\label{eq:H_phi}
H_\phi&=&
  \int_{\varepsilon_{1},\dots,\varepsilon_4} 
\Bigg[ \frac{\phi_1}{\pi} + \frac{\phi_{2\uparrow}}{4\pi} 
\left(\varepsilon_1 + \varepsilon_2 -2 \varepsilon_{0\uparrow}\right) 
\\
\nonumber
&&    + \frac{\phi_{2\downarrow}}{4\pi} 
\left( \varepsilon_3+\varepsilon_4 -2 \varepsilon_{0\downarrow}\right)  \Bigg]
 :\!  b_{\varepsilon_1 \uparrow}^\dagger  b_{\varepsilon_2 \uparrow}  b_{\varepsilon_3 \downarrow}^\dagger  b_{\varepsilon_4 \downarrow} \! : .
\end{eqnarray}
\end{subequations}
It is a perturbative low-energy
expansion involving excitations with respect to a reference
ground state with chemical potentials $\varepsilon_{0\sigma}$ and
distribution function
$n^0_{\varepsilon_{0\sigma}}(\varepsilon) =
\theta(\varepsilon_{0\sigma}-\varepsilon)$.
The dummy reference energies $\varepsilon_{0\sigma}$ should be chosen close
to $\muzerosigma$ for this expansion to make sense.
Here $b^\dagger_{\varepsilon \sigma}$ creates a quasiparticle in a
scattering state with spin $\sigma$ and excitation energy
$\varepsilon - \varepsilon_{0\sigma}$ relative to the reference
state; it already incorporates the zero-temperature phase
shift $\delta_{0\sigma}$. Moreover, $:\,\,:$ denotes normal ordering
w.r.t.\ the reference state, with
\begin{align}
\label{eq:normal-ordering}
: \! b_{\varepsilon \sigma}^\dagger 
b_{\varepsilon \sigma} \! : \, \equiv b_{\varepsilon \sigma}^\dagger 
b_{\varepsilon \sigma} - n^0_{\varepsilon_{0\sigma}} (\varepsilon) \, . 
\end{align}
$H_\alpha$ and $H_\phi$ describe elastic and inelastic scattering
processes, respectively. Their formal structure can be justified using
conformal field theory and symmetry arguments
\cite{affleck1991b,affleck1993,lesage1999a,*lesage1999b}, summarized
in Supplementary Section S-IV of Ref.~\onlinecite{mora2015}.  They
contain the leading and subleading terms in a classification of all
possible perturbations according to their scaling dimensions, which
characterize their importance at low excitation energies with respect
to the reference state. The coupling constants in $\HFL$, together
with the zero-energy phase shifts $\delta_{0\sigma}$, are the model's
nine FL parameters, which we will generically denote by
$\gamma \in \{\delta_{0 \sigma},\alpha_{1\sigma}, \alpha_{2\sigma},
\phi_1, \phi_{2\sigma}\}$.

In the wide-band limit considered here, all FL parameters depend on
the model parameters only in the form
\begin{eqnarray}
\label{eq:invariance-with-shifts}
\gamma = \gamma(U, \Delta, \edsigma - \ezerosigma)  \, , 
\end{eqnarray}
because the chemical potential
$\varepsilon_{0\sigma}$ of our reference ground state
is the only possible point of reference
for the local energies $\varepsilon_{d\sigma}$. Writing
$\varepsilon_{0\sigma} = \varepsilon_0 - \sigma B_0/2$, we thus note
that all FL parameters satisfy the  relations
\begin{eqnarray}
\label{eq:shifting-derivatives}
- \partial_{\varepsilon_0} \gamma =  \partial_{\varepsilon_d} \gamma \, , \qquad
- \partial_{B_0} \gamma =  \partial_B \gamma \, . 
\end{eqnarray}

The form of $\HFL$ in \Eq{eq:H_FL} is similar to that used in
Ref.~\onlinecite{mora2015}, but with two changes, both due to considering
$B\neq 0$.  First, because the magnetic field breaks spin symmetry,
some FL coefficients are now spin-dependent, namely those that occur
in conjunction with excitation energies of the form
$(\varepsilon - \varepsilon_{0\sigma})$.  Second, since the FL theory
of Ref.~\onlinecite{mora2015} was developed around the point $B=0$, the FL
parameters there were taken to be independent of field, and the system's
response to a small field was studied by explicitly including a small
Zeeman term in $\HFL$. In contrast, in the present formulation the FL
parameters are functions of $B$ that explicitly incorporate the full
magnetic-field dependence of all ground state properties, hence our
$\HFL$ does not need an explicit Zeeman term.

To conclude this subsection, we note that the form of $\HFL$ presented
above can be derived by an explicit calculation in a particular
limiting case: the Kondo limit of the Anderson Hamiltonian where it
can be mapped onto the Kondo Hamiltonian, studied in the limit of very
large magnetic field. By doing perturbation theory in the spin-flip
terms of the Kondo Hamiltonian, one arrives at effective interaction
terms that have precisely the form of $H_\alpha$ and $H_\phi$
above. This calculation, presented in detail in
Appendix~\ref{appen:kondo}, is highly instructive, because it
elucidates very clearly how the reference energies
$\varepsilon_{0\sigma}$ enter the analysis and how the relations
\eqref{eq:invariance-with-shifts} and \eqref{eq:shifting-derivatives}
come about.

\subsection{Relating FL parameters to local observables} 

Having presented the general form of $\HFL$, the next step is to
express the FL parameters in terms of ground state observables. The
corresponding relations are conveniently derived by examining the
elastic phase shift of a single quasiparticle
excitation~\cite{mora2009a}.  We suppose that the system is in an
arbitrary state, not too far from the ground state, characterized by
the spin-dependent number distribution
$\langle b_{\varepsilon \sigma}^\dagger b_{\varepsilon' \sigma}
\rangle = n_{\sigma} (\varepsilon) \delta(\varepsilon-\varepsilon')$,
with arbitrary $n_{\sigma} (\varepsilon)$.  The elastic phase shift of
a quasiparticle with energy $\varepsilon$ and spin $\sigma$ scattered
off this state is obtained from the elastic part $H_\alpha$, in
addition to the Hartree diagrams inherited from $H_\phi$, thus
$\delta_\sigma (\varepsilon,n_{\sigma'}) = \delta_{0\sigma} - \pi \partial \langle
H_\alpha + H_\phi \rangle/\partial_{n_\sigma (\varepsilon)}$.
One finds the expansion
\begin{align}
\label{pshift-Anderson-finite-B}
& \delta_{\sigma}  (\varepsilon,n_{\sigma'}) =
    \dzerosigma +  \alpha_{1\sigma} (\varepsilon-\varepsilon_{0\sigma})  +
  \alpha_{2\sigma}  (\varepsilon-\varepsilon_{0\sigma})^2 
  \\[1mm] 
  \nonumber & \quad - \int_{\varepsilon'} 
\left[\phi_1 + \tfrac{1}{2} \phi_{2\sigma} (\varepsilon - \varepsilon_{0\sigma}) 
+ \tfrac{1}{2} \phi_{2\bar \sigma} (\varepsilon' - \varepsilon_{0\bar \sigma}) 
\right] \delta n_{\bar \sigma}
              (\varepsilon') \, . 
\end{align}
Due to the normal ordering prescription for $H_\phi$, all terms
stemming from the latter involve the difference between the actual and
reference distribution functions,
$\delta n_{\bar{\sigma}} = n_{\bar{\sigma}} -
n^0_{\varepsilon_{0\bar\sigma}} $,
where $\bar{\sigma}$ denotes the spin opposite to $\sigma$.  Now,
though expansion (\ref{pshift-Anderson-finite-B}) depends on
$\ezerosigma$ both explicitly and via the FL parameters
$\gamma(U,\Delta,\edsigma - \ezerosigma)$, these dependencies have to
conspire in such a way that the phase shift is actually
\textit{independent} of $\ezerosigma$.  Thus the following conditions
must be satisfied:
\begin{align}
\label{eq:floating-condition-I}
\partial_{\ezero} \delta_{\sigma}  (\varepsilon,n_{\sigma'}) = 0 , 
\qquad 
\partial_{\Bzero} \delta_{\sigma}  (\varepsilon,n_{\sigma'}) = 0 .
\end{align} 
Inserting Eq.~\eqref{pshift-Anderson-finite-B}, setting the coefficients 
of the various terms in the expansion (const., $\sim (\varepsilon-
\ezerosigma), \propto \int \delta n_{\bar \sigma}$) to zero
and exploiting \Eqs{eq:shifting-derivatives}, we obtain 
a set of linear relations among the FL parameters, 
to be called ``Fermi liquid relations'': 
\begin{subequations}
\label{eq:FL-relations-I}
\begin{align}
\frac{\partial \dzerosigma}{\partial \ed}
 & = \phi_1 - \alpha_{1\sigma} , 
& 
\frac{\partial \dzerosigma}{\partial B} 
& = \tfrac{\sigma}{2}(\phi_1 + \alpha_{1\sigma}) , 
\\
\frac{\partial \alpha_{1\sigma}}{\partial \ed}  
& = \tfrac{1}{2} \phi_{2 \sigma}
- 2 \alpha_{2\sigma}  ,  \hspace{-3mm}
&
\frac{\partial \alpha_{1\sigma}}{\partial B} 
& = \tfrac{\sigma}{2}(\tfrac{1}{2} \phi_{2 \sigma}
+ 2 \alpha_{2\sigma})  ,  
\\
\frac{\partial \phi_1}{\partial \ed}
& = - \tfrac{1}{2} (\phi_{2 \uparrow} + \phi_{2 \downarrow})  , 
\hspace{-2mm}
&
\frac{\partial \phi_1}{\partial B}
& =  \tfrac{1}{4} (\phi_{2 \uparrow} - \phi_{2 \downarrow})  . 
\end{align}
\end{subequations}
They are important for three reasons.  First, for fixed values of the
model parameters, they ensure by construction that
$\delta_{\sigma} (\varepsilon , n_{\sigma'})$ is invariant under
spin-dependent shifts of the dummy reference energies
$\ezerosigma$. Second, for fixed values of $\ezerosigma$, they ensure
that for any distribution $n_{\sigma'}$ with well-defined chemical
potentials $\mu_{\sigma'}$, the function
$\delta_{\sigma} (\varepsilon , n_{\sigma'})$ is invariant, up to a
shift in $\varepsilon$, under simultaneous spin-dependent shifts [cf.\
Eq.~\eqref{eq:general-shifts}] of the physical model parameters
$\varepsilon_{d\sigma'}$ and $\mu_{\sigma'}$, say by
$\delta \mu_{\sigma'} = \delta \mu - \tfrac{1}{2} \sigma' h$:
\begin{eqnarray}
\label{eq:physical-shifts-for-phase}
  \delta_{\sigma}  (\varepsilon \!+\! \delta \mu_\sigma, n_{\sigma'})
  \Big|_{\varepsilon_{d\sigma^\prime} 
+ \delta \mu_{\sigma'} ,\mu_{\sigma'} + \delta \mu_{\sigma'}} \!\! \!\!
=   \delta_{\sigma}  (\varepsilon,n_{\sigma'})
\Big|_{\varepsilon_{d \sigma'},\mu_{\sigma'}} \!\!  . \qquad
  \phantom{.}
\end{eqnarray} 
Conversely, an alternative way to derive
Eqs.~\eqref{eq:FL-relations-I} is to impose
\Eq{eq:physical-shifts-for-phase} as a condition on the expansion
(\ref{pshift-Anderson-finite-B}).  [Verifying this is particularly
simple at zero temperature, e.g.\ using
$n_{\sigma'} = n^0_{\muzerosigma}$.] In the parlance of Nozi\`eres
\cite{nozieres1974,*nozieres1974b,*nozieres1978},
Eq.~\eqref{eq:physical-shifts-for-phase} is the ``strong
universality'' version of his ``floating Kondo resonance'' argument,
applied to the Anderson model. Pictorially speaking, for each spin
species the phase shift function ``floats'' on the Fermi sea of
corresponding spin: if the Fermi surface $\mu_\sigma$ and local level
$\edsigma$ for spin $\sigma$ are both shifted by $\delta \mu_\sigma$,
the phase shift function $\delta_\sigma(\varepsilon,n_{\sigma'})$
shifts along without changing its shape.

Third, the Fermi liquid relations, in conjunction with Friedel's sum
rule, can be used to link the FL parameters to ground state values of
local observables.  To this end, we henceforth set
$\ezerosigma = \muzerosigma$ and focus on the case of zero temperature
with ground state distribution $\nzero_\muzerosigma$. Then only the
first term in Eq.~\eqref{pshift-Anderson-finite-B} survives when
writing down Friedel's sum rule for the phase shift at the chemical
potential:
\begin{equation}
\label{friedel2}
\delta_{\sigma}  (\muzerosigma,\nzero_\muzerosigmaprime) = \dzerosigma = 
\pi n_{d\sigma} \, . 
\end{equation}
Let 
  $n_d = \sum_\sigma n_{d\sigma}$ and
  $m_d = \frac{1}{2} \sum_\sigma \sigma n_{d\sigma} $ denote the
  average local charge and magnetization, respectively, 
and let us introduce corresponding even and odd
linear combinations of the spin-dependent FL parameters, to be denoted
without or with overbars, e.g.\
$\alpha_1 = \frac{1}{2} \sum_\sigma \alpha_{1\sigma}$ and
$\overline \alpha_1 = \frac{1}{2} \sum_\sigma \sigma \alpha_{1\sigma}$.
Then we have $n_d = 2\delta_0/\pi$ and $m_d = \overline \delta_0/\pi$.  By
differentiating these relations with respect to $\ed$ and $B$ we
obtain various local susceptibilities, which can be expressed, 
via the derivates occurring in Eq.~\eqref{eq:FL-relations-I}, 
as linear combinations of FL parameters: 
\begin{subequations}
\label{eq:susceptibilities-largeB}
\begin{alignat}{3}
\label{FLiden2-largeB}
\chi_c  = & & 
- \frac{\partial \nd}{\partial \ed} \;\, 
& =  
\; \, \frac{2}{\pi} (\alpha_1 - \phi_1 ) \, ,   
\\
\label{spinsus-largeB}
\chi_s = &   &  \frac{\partial \md}{\partial B}  \;
&  = 
\frac{1}{2 \pi} (\alpha_1 + \phi_1) \, , 
\\
  \label{spinsus-mixe-largeB}
  \chi_m =  & & - \frac{\partial \md}{\partial \ed} \;
 & =  \frac{\partial \nd}{\partial B}  = \frac{\balpha_1}{\pi} \, , 
\\
 \frac{\partial \chi_c}{\partial \ed}  =  & &
 - \frac{\partial^2 \nd}{\partial \ed^2} 
& = - \frac{4}{\pi} \left[ \alpha_2  -\frac{3}{4}\phi_2 \right]  , \quad 
\phantom{.}
\\
\frac{\partial \chi_s}{\partial B} = &  &  \phantom{-}  
\frac{\partial^2 \md}{\partial B^2} 
& = \, \frac{1}{2\pi}  \left[ \balpha_2  +\frac{3}{4} \,  \bphi_2 \right]  , 
\\
\label{eq:dchicdB}
 \frac{\partial \chi_m}{\partial \ed } = & & 
\; - \frac{\partial^2 m_d}{\partial \ed^2} 
&  =   \phantom{-} \frac{\partial^2 n_d }{\partial B \partial \ed}
= - \frac{2}{\pi}  \left[ \balpha_2  - \frac{\bphi_2}{4}  \right]  ,
\\
\frac{\partial   \chi_m}{\partial B} = & &
\frac{\partial^2 \nd}{\partial B^2} 
& =  - \frac{\partial^2 \md}{\partial \ed \partial B} 
= \phantom{-}\frac{1}{\pi} \left[ \alpha_2  +\frac{\phi_2}{4} \right]  .  
\end{alignat}
\end{subequations}
\Eq{spinsus-mixe-largeB} reproduces a standard thermodynamic identity,
and implies similar identities for higher derivates,
$\partial \chi_{m}/\partial \ed = -
\partial \chi_{c}/\partial B$ and 
$\partial \chi_{m}/\partial B = -
\partial \chi_{s}/\partial \ed$.
By inverting the above relations, we obtain
the FL parameters in terms of local ground state 
susceptibilities: 
\begin{subequations}
\label{fermic}
\begin{align}
\frac{\alpha_1}\pi  & =  \chi_s + \frac{1}{4} \chi_c \, , &  
\frac{\alpha_2}\pi  & = \frac{3}{4} \frac{\partial \chi_m}{\partial B}
 - \frac{1}{16}\frac{\partial \chi_c}{\partial \ed} \ , 
\\
\frac {\phi_1}\pi &  =  \chi_s - \frac{1}{4} \chi_c , & 
\frac{\phi_2} \pi & =    \phantom{\frac{3}{4}}
\frac{\partial \chi_m}{\partial B} + 
\frac{1}{4}\frac{\partial \chi_c}{\partial \ed} \, , 
\\
\frac{\balpha_1}\pi   & =  \chi_m, & 
\frac{\balpha_2}\pi   &   = \frac{1}{2} \frac{\partial \chi_s}{\partial B} 
- \frac{3}{8}  \frac{\partial \chi_m}{\partial \ed} \, ,  
\\ 
& & \frac{\bphi_2}\pi    & = \, 2 \frac{\partial \chi_s}{\partial B} 
+ \frac{1}{2} \frac{\partial \chi_m}{\partial \ed} \,  ,
\end{align}
\end{subequations}
implying that $\phi_2 = - \partial_{\ed} \phi_1$ and
$\bphi_2 = 2 \partial_B \phi_1$.  These equations are a central
technical result of this paper.  Those for the even FL parameters
$\alpha_{1,2}$ and $\phi_{1,2}$ are equivalent to the ones obtained,
for zero field, in Ref.~\onlinecite{mora2015}. 
The expressions for $\alpha_1$ and $\phi_1$ have been shown~\cite{mora2015} to be equivalent to
the relation
\begin{eqnarray}
  \label{eq:4chis+chic}
  \frac{4 \chi_s}{(g \mu_B)^2} + \chi_c =
\frac{6 \gamma_\imp}{\pi^2 k_B^2} \; ,
\end{eqnarray}
(physical units have been reinstated in this equation) between the
spin/charge susceptibilities and the impurity specific heat
coefficient $\gamma_\imp$~\cite{hewson1993}.  This relation in fact
derives from Ward
identities~\cite{yosida1970,*yamada1975a,*yosida1975,*yamada1975b,yoshimori1976}
associated with the $U(1)$ symmetry of the model.  We expect that the
other expressions in Eqs.~\eqref{fermic} also originate from Ward
identities involving higher-order derivatives.

Equations \eqref{fermic} can be checked independently in two limits:
for a non-interacting impurity, and at large magnetic field in the
Kondo model, see Appendix~\ref{appen:kondo} for the latter. The former
case, $U=0$, reduces to a resonant level model in which spin and
charge susceptibilities are easily obtained.  We have verified that
they give $\phi_1 = \phi_{2 \sigma} =0$, so that the interaction
$H_\phi =0$ in Eq.~\eqref{eq:H_FL} vanishes, and that the phase shift
expansion~\eqref{pshift-Anderson-finite-B} reproduces that expected
for the resonant level model.

  All of the susceptibilities introduced above are calculable exactly
  by Bethe Ansatz, and hence the same is true for all the FL
  parameters. In the particle-hole symmetric case,
  $\varepsilon_d = -U/2$, semi-analytical expressions for the local
  charge and magnetization have been derived with the help of the
  Wiener-Hopf method. A comprehensive review on this approach can be
  found in Ref.~\onlinecite{tsvelick1983} and we summarize the
  resulting analytical expressions in the Supplemental
  Material~\footnote{See Supplemental Material, where the Bethe Ansatz
    expressions used here are reviewed.}. They have been used to
  produce Fig.~\ref{fig-kondocA} to \ref{fig-kondocv} below with excellent
  accuracy.

  Away from particle-hole symmetry, where the Wiener-Hopf method is
  not applicable, the Bethe Ansatz coupled integral equations (see
  Eq.~(S3a) and~(S3b) in the Supplemental Material~\cite{Note1}) have
  to be solved numerically. This direct method is used in
  Figs.~\ref{fig:cvdens} and~\ref{fig:cvcuts}. In
  Fig.~\ref{fig-kondocv} we have verified that at particle-hole
  symmetry it agrees nicely with the accurate Wiener-Hopf
  solution.

To conclude this subsection, 
we briefly discuss some special cases, for future reference:

(i) \textit{Zero magnetic field:} 
Eqs.~\eqref{fermic} for the odd FL parameters yield
zero for $B=0$, 
\begin{eqnarray}
  \label{eq:odd-zeroB}
  \balpha_1 = \balpha_2 = \bphi_2 = 0 \, , 
\end{eqnarray}
since $m_d$ is an antisymmetric
function of $B$. 

(ii) \textit{Particle-hole symmetry:}
At $\ed = -U/2$ we have
\begin{equation}
\label{eq:particle-hole-symmetry}
n_d = 1, \;\; \delta_{0\sigma} = \pi \left( \tfrac1 2+ \sigma m_d \right) , 
\;\; \balpha_1 = \alpha_2 = \phi_2 = 0 , 
\end{equation}
for any $B$. The three FL parameters vanish since $n_d-1$ is an
antisymmetric function of $\ed-U/2$, implying the same for 
$\chi_m$ and $\partial \chi_c/\partial \ed$, so that both vanish at
$\ed = -U/2$.

(iii) \textit{Kondo limit:} If the limit $U/\Delta \to \infty$ is
taken at particle-hole symmetry while maintaining a finite Kondo
temperature, local charge fluctuations are frozen out completely and
the Anderson model maps onto the Kondo model. All susceptibilities
involving derivatives of $n_d$ with respect to $\ed$ vanish,
namely
  $\chi_c = \chi_m = \partial_{\ed} \chi_c =\partial_{\ed} \chi_s
  = \partial_B \chi_c = 0$,
so that \Eqs{eq:particle-hole-symmetry} are supplemented by
\begin{equation}\label{relations-kondo}
\frac{\alpha_1}{\pi} = \frac{\phi_1}{\pi} = 
\chi_s, 
\qquad \frac{4 \balpha_2}{\pi} = \frac{\bphi_2}{\pi} = 2 
\frac{\partial \chi_s}{\partial B} \, . 
\end{equation}
Since $\chi_s$ and $\partial \chi_s/\partial B$
are strictly positive and negative, respectively, 
the same is true for $\alpha_1,\phi_1$ and $\balpha_2,\bphi_2$. 

(iv) \textit{Kondo limit at large fields:} 
In the limit $B \gg \Tk$  of the Kondo model, its
Bethe Ansatz solution yields the following results for the leading
asymptotic behavior of the magnetization and its derivates,
with $\beta_r = \frac{\pi}{8}(B/\Tk)^2$: 
\begin{subequations}
  \label{eq:large-B}
\begin{align}
  m_d & = \frac{1}{2} - \frac{1}{2\ln \beta_r} , 
\quad 
\chi_s = \frac{1}{B \, (\ln \beta_r)^2} , 
\\
\frac{\partial \chi_s}{\partial B}
&  = - \frac{1}{B^2 \, (\ln \beta_r)^2} .
\end{align}
\end{subequations}
Thus, all the FL parameters in \Eq{relations-kondo} vanish
asymptotically in the large-field limit.

\subsection{Characteristic FL energy scale}
\label{sec:Estar}

As mentioned repeatedly above, the FL approach only holds for
excitation energies sufficiently small, say
$|\varepsilon - \muzerosigma| \ll \Estar$, that all terms in expansion
\eqref{pshift-Anderson-finite-B} for
$\delta_{\sigma} (\varepsilon,n_{\sigma'}) - \dzerosigma$ are
small. In the local moment regime of the Anderson model, the FL scale
$\Estar$ can be associated with the Kondo temperature $\Tk$, but in
the present context we need a definition applicable in the full
parameter space of the Anderson model.  Following
Ref.~\onlinecite{mora2015}, we define $\Estar$ in terms
of the FL coefficient of the leading term in expansion
\eqref{pshift-Anderson-finite-B},
\begin{eqnarray}
  \label{eq:define-Estar}
 \Estar = \frac{\pi}{4\alpha_1 } = \frac{1}{4 \chi_s + \chi_c} \, , 
\end{eqnarray}
and $\Tk$ in terms of the zero-field
spin susceptibility, 
\begin{equation}\label{eq:tksus}
T_K  = \frac{1}{4 \chi_s^{B=0}} \, .
\end{equation}
While both definitions involve some arbitrariness, they are mutually
consistent, in that the zero-field value of $\Estar$ equals $\Tk$
in the Kondo limit $U/\Delta \to \infty$, where we have
\begin{align}
  \label{eq:EstarKondolimit}
  \Estar^{B=0} = \Tk, \qquad 
\Estar^{B\gg \Tk} = \tfrac{1}{4} B (\ln \beta_r)^2 \, .
\end{align}
More generally, $\Estar^{B=0}$ and $\Tk$ are roughly equal throughout
the local-moment regime where $\chi_c \simeq 0$, i.e.\ for
$U\gg \Delta$ and $-U + \Delta \lesssim \ed \lesssim - \Delta$. In this regime,
$\Tk$ is well described by the analytic formula (af)
\cite{haldane1978,haldane1978a,mora2015}
\begin{align}
\label{eq:estar}
\Tkaf = \sqrt{\tfrac{U\Delta}2}\, 
e^{\pi\left[\frac{\Delta}{2U} - \frac{U}{8 \Delta}\right]} e^{x^2} \, , \quad
\end{align}
where $x = (\ed + U/2)\sqrt{\pi/(2 \Delta U)}$ measures the distance
to the particle-hole symmetric point.  At the latter, 
$\Tkaf|_{x=0}$ can be derived analytically  from the Bethe-Ansatz
equations for $\chi_s^{B=0}$ \cite{tsvelick1983}. The factor
$e^{x^2}$, familiar from Haldane's RG treatment of the Anderson model
\cite{haldane1978}, phenomenologically includes the effect
of particle-hole asymmetry. Throughout the local
moment regime, Eq.~\eqref{eq:estar}  yields excellent
agreement with a direct numerical evaluation of \Eq{eq:tksus} via 
the Bethe-Ansatz equations for $\chi^{B=0}_s$ (see Fig.~\ref{fig:cvdens} below).

\section{Spectral function and non-linear conductance}
\label{sec:spectral-function-conductance}

\subsection{General results}
\label{sec:A-G-general}
 
For the remainder of this paper we consider a single-level quantum dot
with symmetric tunnel couplings to left and right leads with chemical
potentials $\pm eV/2$, described by the two-lead, single-level
Anderson model. The non-linear conductance of this system can be
expressed by the Meir-Wingreen formula as \cite{meir92}
\begin{eqnarray}
  \label{eq:Meir-Wingreen}
  G (V,T) = \partial_V \frac{e}{h}  
  \int_\ve   \left[f_L(\ve) - f_R(\ve) \right] A (\ve) \, . 
\end{eqnarray}
Here $f_{L/R} (\ve) = [e^{(\ve\mp eV/2)/T}\!+\!1 ]^{-1}$ are the
distribution functions of the left and right leads,
$A(\varepsilon) = \sum_\sigma A_\sigma (\varepsilon)$ is the local
spectral function with spin components
$A_\sigma (\varepsilon) = - \pi \nu_0 \, \text{Im}
\mathcal{T}_\sigma(\varepsilon)$,
and $\mathcal{T}_\sigma (\varepsilon)$ is the $T$-matrix for spin
$\sigma$ conduction electrons scattering off the local level.  A FL
calculation of the low-energy behavior of $A_\sigma(\ve)$ and $G(V,T)$
has been performed in detail at zero magnetic field in
Ref.~\onlinecite{mora2015}, following similar studies in
Refs.~\onlinecite{mora2008,vitushinsky2008,mora2009,horig2014}. The
strategy of the calculation is rather straightforward. First one
introduces even and odd linear combinations of operators from the two
leads. The odd ones decouple, resulting in an effective one-lead
Anderson model for a dot coupled to the even lead, whose low-energy
behavior is described by the Hamiltonian $\HFL$ introduced
above. Then, in the spirit of the standard Landauer-B\"uttiker
formalism~\cite{blanter2000}, the current operator is expanded over a
convenient single-particle basis of scattering states accounting for
both the lead-dot geometry and the FL elastic phase shifts.
Interactions between electrons stemming from $H_\phi$ are included
perturbatively when calculating the average current in the Keldysh
formalism~\cite{kamenev2009}.

The calculation described above trivially generalizes to the case of
nonzero field, since the two spin components give separate
contributions to the current. 
The results from Ref.~\onlinecite{mora2015} for the low-energy
expansion of the conductance can thus be directly taken over,
modified merely by supplying FL parameters with spin indices.
A corresponding low-energy expansion for the spectral function can 
then be deduced via \Eq{eq:Meir-Wingreen}. We now present
the results obtained in this manner, starting with the $T$-matrix
and spectral function, since these form the basis for understanding
the resulting physical behavior.

For the $T$-matrix, written as the sum of elastic and
inelastic contributions, the results of Ref.~\cite{mora2015} (Supplementary
Section S-V) imply 
\begin{subequations}
\begin{align}
  \label{eq:T-elastic}
   {\cal T}^{\rm el}_\sigma (\varepsilon) 
& = - \frac{i}{2 \pi \nu_0} \left( 1 - e^{2 i \delta_\sigma (\varepsilon)} \right),
\\
  \label{eq:T-inelastic}
{\cal T}^{\rm inel}_\sigma (\varepsilon) & 
= - \frac{i e^{2 i \dzerosigma}}{2 \pi \nu_0} \phi_1^2 
\left[ \varepsilon^2 + (\pi T)^2 + \tfrac{3}{4}(eV)^2\right] .
\end{align}
\end{subequations}
Here $  {\cal T}^{\rm el}_\sigma (\varepsilon)$
is determined by the phase shift
$\delta_\sigma(\ve)$ obtained from \Eq{pshift-Anderson-finite-B} using
$n_\sigma(\ve) = \frac{1}{2}\left[f_L(\ve) + f_R(\ve) \right]$ as
quasiparticle distribution function for the even lead:
\begin{flalign}
  \label{eq:phase-shift-two-lead-dot}
& \delta_\sigma (\varepsilon) 
= \dzerosigma   \! + \! \alpha_{1\sigma} \varepsilon \! + \! 
\alpha_{2\sigma}  \varepsilon^2 
\! - \!  
\tfrac{1}{12} \phi_{2\bar \sigma} \! \left[(\pi T)^2 \!+\! 
\tfrac{3}{4}(e V)^2\right]\! . \hspace{-1cm} & 
\end{flalign} (In Ref.~\cite{Filippone2017} this equation
  contained an error, which lead to incorrect physical conclusions,
  see Ref.~\footnote{ In Ref.~\cite{Filippone2017}, the last
      term in Eq.~\eqref{eq:phase-shift-two-lead-dot} erroneously
      contained a factor $\phi_{2 \sigma}$ instead of
      $\phi_{2 \bar \sigma}$.  Consequently most subsequent
      expressions for $\tilde C_A$ were incorrect. For example,
      Eq.~\eqref{eq:tildeCA-main-paper} contained an $\phi_{2 \sigma}$
      instead of $\phi_{2 \bar \sigma}$, and
      Eq.~\eqref{eq:tildec_A-prediction-particle-hole} an incorrect
      sign, minus instead of plus, in front of the $\overline \phi_2$
      term. This lead to the wrong conclusion that in the Kondo limit
      $\tilde C_A$ and $C_V$ remain positive for large fields --
      insteady, they turn negative; and that ``the conductance retains
      a (very weak) zero-bias maximum even for strong magnetic
      fields'' -- instead, the zero-bias maximum turns into a minimum
      at a field $B_V$ of order $\Tk$.}.) Note that the inelastic
$T$-matrix has the same dependence on temperature and bias, which
occur only in the combination $(\pi T)^2 \! + \!  \frac{3}{4}(e V)^2$
\footnote{Expression \eqref{eq:T-inelastic} for
  ${\cal T}_\sigma^{\rm inel} $ stems from the quadratic contribution
  of the $\phi_1$ term in $H_\phi$ to the quasi-particle
  self-energy. Its $\ve^2 + (\pi T)^2$ contribution is well known in
  an equilibrium context
  \cite{nozieres1974,*nozieres1974b,*nozieres1978,affleck1993}.  We
  obtained its $(eV)^2$ contribution as follows. Since $\ve^2$, $T^2$
  and $(eV)^2$ all characterize the phase space available for
  inelastically scattering a quasiparticle having energy $\ve$, their
  contributions all have the same general form, differing only by
  numerical prefactors.  We deduced that of $(eV)^2$ to be
  $\frac{3}{4}$ by inserting a general low-energy expansion for
  $A(\ve)$ into \eqref{eq:Meir-Wingreen} for $G(T,V)$, expanding the
  latter in the form \eqref{eq:c's-B-dependent} and equating the
  resulting expression for $\CV$ to the quantity $c_V/\Estar^2 $ found
  in Ref.~\cite{mora2015} by a direct calculation of the current. The
  resulting combination
  $\left[ \varepsilon^2 + (\pi T)^2 + \tfrac{3}{4}(eV)^2\right]$ in
  \Eq{eq:T-inelastic} for ${\cal T}^{\rm inel}_\sigma (\varepsilon)$
  is consistent with that reported in Eq.~(43) of
  Ref.~\cite{Oguri2005b} for the imaginary part of the local
  self-energy in renormalized perturbation theory.}.
This is significant, since it implies that knowing the spectral
  function's leading temperature dependence in \textit{equilibrium}
  suffices to deduce its leading bias dependence in nonequilibrium.
The spectral function, expanded to second order in $\ve$, $T$ and
$eV$, can thus be written as \footnote{The numerical prefactor for
  $\protect\tCA$ was chosen to ensure that
  $\protect\tCA = \protect\CA$ in the Kondo limit at zero field, see
  Eq.~(\ref{eq:CA-tCA-Kondo-limit-B=0}).}
  \begin{flalign}  
\label{eq:expandAepsilon-general}
& A (\varepsilon)  =   A_0  +  A_1 \ve 
  - \tCA \! \left[\tfrac{1}{3} (\pi T)^2 \! + \!  
 \tfrac{1}{4}(e V)^2\right] -  \CA \ve^2 , \hspace{-1cm} & 
\end{flalign}
with expansion coefficients: 
\begin{subequations}
\label{subeq:Acoefficients} 
\begin{flalign}
A_0 & =  \sum_\sigma \sin^2(\dzerosigma) , \quad 
A_1 = \sum_\sigma \alpha_{1\sigma} \sin(2 \dzerosigma), 
 \\
\label{eq:tildeCA-main-paper}
\tCA & = 
-  \sum_\sigma
\left[ 
\tfrac{3}{2} \phi_1^2 \cos(2 \dzerosigma) - 
 \tfrac{1}{4} \phi_{2  \bar \sigma} \sin(2 \dzerosigma) \right] \! , 
\\
\label{eq:CA-main-paper}
\CA & = -  \sum_\sigma \! \left[
(\alpha_{1 \sigma}^2 \!+\! \tfrac{1}{2} \phi_1^2)  \cos(2 \dzerosigma)
+  \alpha_{2 \sigma} \sin(2 \dzerosigma) \right] \! . \hspace{-1cm} &
 \end{flalign}  
\end{subequations}
These results hold for all values of $U$, $\Delta$,
$\varepsilon_d$ and $B$.

Inserting \Eq{eq:expandAepsilon-general}
into \eqref{eq:Meir-Wingreen} and using the relations
\begin{align}
  \label{eq:explain-e2-vs-T2-V2}
&  \partial_V \!\!
 \int_\ve    \left[f_L \!-\! f_R \right]  \ve^2
=  e\left[ \tfrac{1}{3} (\pi T)^2 \!+\! \tfrac{1}{4}(e V)^2 \right] , 
\\
& \nonumber
 \partial_V \!\! 
 \int_\ve 
 \left[f_L \!-\! f_R \right] \!
\left[ \tfrac{1}{3}(\pi T)^2 \!+\! \tfrac{1}{4}(e V)^2 \right] 
=   e \! \left[ \tfrac{1}{3}(\pi T)^2 \!+\! \tfrac{3}{4}(e V)^2\right]\!, 
\end{align}
one obtains an expansion for the conductance of the form 
\begin{align} 
\label{eq:c's-B-dependent}
  G (V,T) & = \tG - 
  (2e^2/h) \left[ \CT T^2 + \CV (e V)^2 \right] .
\end{align}
Here $\tG = \frac{1}{2} A_0 G_0$ is the zero-temperature, linear
conductance, $G_0 = 2e^2/h$ is the conductance
quantum, and the expansion coefficients of the quadratic terms are
\begin{align}
\label{eq:cT-cV-cA-ctildeA} 
\CT  & = \tfrac{1}{6}\pi^2 (\tCA + \CA) \, , 
\quad \CV = \tfrac{3}{8}(\tCA + \tfrac{1}{3} \CA ) \, . 
\end{align}
Eqs.~\eqref{eq:cT-cV-cA-ctildeA} and \eqref{subeq:Acoefficients}
  are consistent with expressions for $C_T$ and $C_V$ recently derived
  by Oguri and Hewson \cite{Oguri2017,*Oguri2017a,*Oguri2017b} (see
  Appendix~\ref{app:oh}).

The four $C$ coefficients introduced above
all have dimensions of (energy)$^{-2}$.  If we express them as
\begin{eqnarray}
\label{eq:cCEstar}
\tCA = \frac{\tilde c_A}{\Estar^2}, \quad  
C_X =  \frac{c_X}{\Estar^2} , \quad X = A,V,T \, , 
\end{eqnarray}
where $\Estar$ is the FL scale of \Eq{eq:define-Estar}, the resulting
four $c$ coefficients are dimensionless, with $c_T$ and $c_V$
corresponding to the coefficients calculated in
Ref.~\onlinecite{mora2015}. For asymmetric couplings to the
  leads~\cite{aligia2012,munoz2013,aligia2014}, not considered
    here, the conductance also contains a term linear in $V$, as also
  discussed in Ref.~\cite{mora2009}, where the same formalism has been
  applied.

  Equations \eqref{subeq:Acoefficients} instructively reveal which
  role the various FL parameters play in determining the shape of the
  local spectral function $A(\ve)$ at the chemical potential,
  characterized by its ``height'' $A(0)$, slope $A_1$ and curvature
  $\CA$.  The ground state phase shifts
  $\dzerosigma$ fix the height at zero temperature and bias,
  $A_0$. The elastic couplings $\alpha_{1\sigma}$ and
  $\alpha_{2\sigma}$ of $H_\alpha$ affect only the slope and
  curvature, but not the height.  The inelastic couplings $\phi_1 $
  and $\phi_{2\sigma}$ of $H_\phi$ determine the leading effect of
  temperature and bias on the height via $\tCA$, while $\phi_1$ also
  contributes to the curvature $\CA$.  Moreover, via the sine and
  cosine factors the relative contributions of all terms depend
  sensitively on the ground state phase shifts $\dzerosigma$, and
  hence can change significantly when these are tuned via changing
  parameters such as $B$ or $\ed$.

\subsection{Spectral function at particle-hole symmetry}
\label{sec:A-particle-hole}

When the single-level, two-lead Anderson model is tuned into the local
moment regime, the local spectral function exhibits a Kondo peak that
splits with magnetic field.  Correspondingly, the non-linear
conductance exhibits a zero-bias peak that likewise
splits with increasing field. Our goal is to use
FL theory to study the peak splittings of both the spectral function
and the nonlinear conductance in quantitative detail. For this
purpose, we will focus on the particle-hole symmetric point
in this subsection and the next, leaving
particle-hole asymmetry to Subsection~\ref{sec:oph}.

We begin with a qualitative discussion, based on the results of
numerous previous studies of the local moment regime
\cite{costi2000,wright2011,zitko2009,moore2000,glazman2005,rosch2003,rosch2005,weichselbaum2009}. At
zero field, the two components of the local spectral function,
$A_\uparrow$ and $A_\downarrow$, both exhibit a Kondo peak at zero
energy.  An increasing field weakens these peaks and shifts them in
opposite directions. When their splitting exceeds their width, which
happens for $B$ of order $\Tk$, then $A = A_\uparrow + A_\downarrow$
develops a local minimum at zero energy, implying that $\CA$ changes
from positive to negative. We will denote the ``splitting field''
where $\CA=0$ by $B_A$.  [For $B \gg T_K$ the subpeaks in
$A_{\uparrow,\downarrow}$ are located at $\varepsilon \simeq \pm B$,
modulo corrections of order $\mp B/\log(B/\Tk)$
\cite{moore2000,rosch2003,weichselbaum2009}.]  

For small fields, where the Kondo peak is well developed, an
  increasing temperature or bias tends to weaken it, thus reducing the
  zero-energy spectral height $A(0)$. We thus expect $\tCA$ to be a
  decreasing but positive function of $B$ for small fields. However,
  this trend can be expected to be reversed for fields of order $\Tk$
  or larger, where Kondo correlations are weak or absent, in which
  case we may expect $\tCA$ to become negative. We will denote this
  field by $\tBA$.

\begin{figure}
\includegraphics[width=0.94\columnwidth]{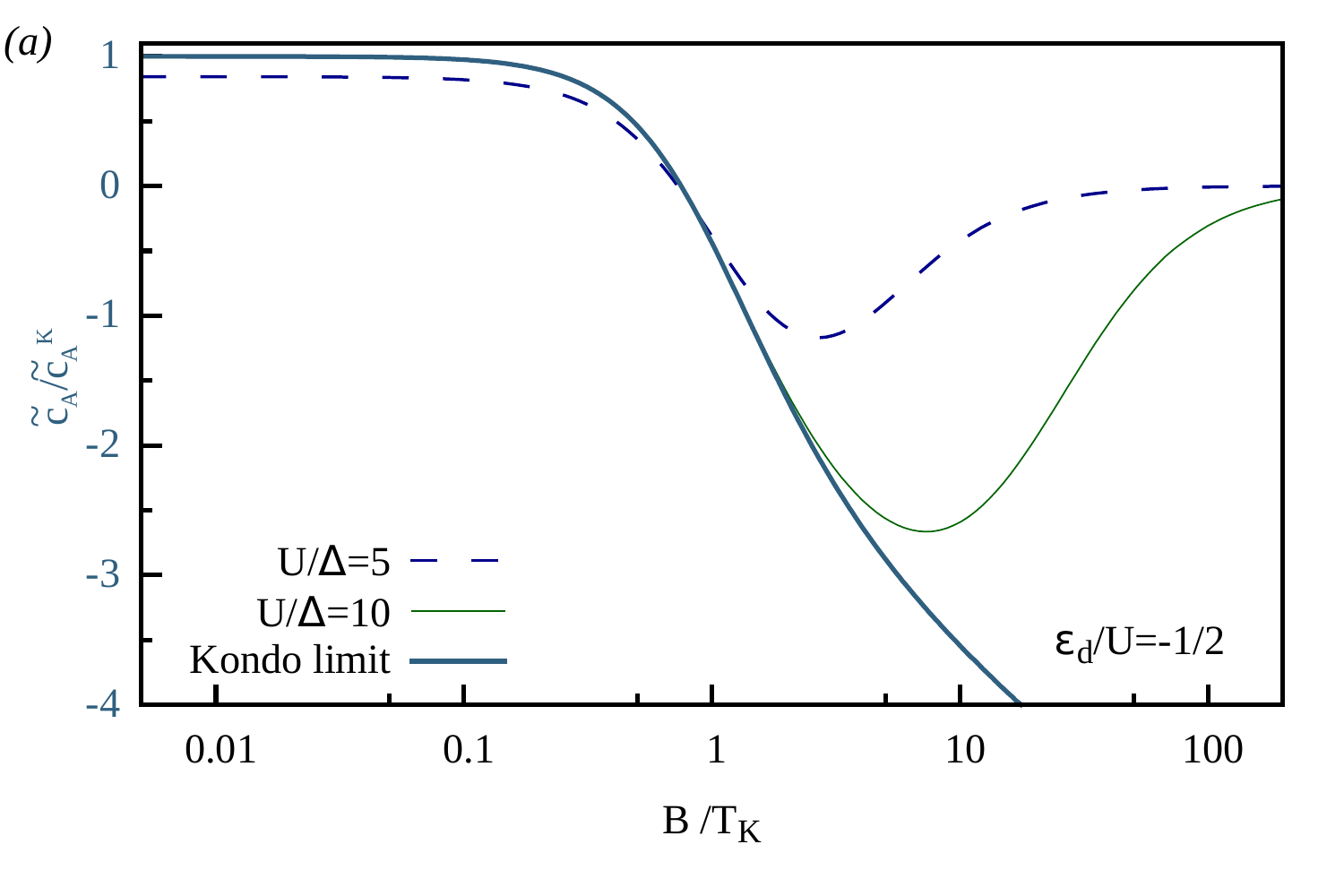}
\includegraphics[width=\columnwidth]{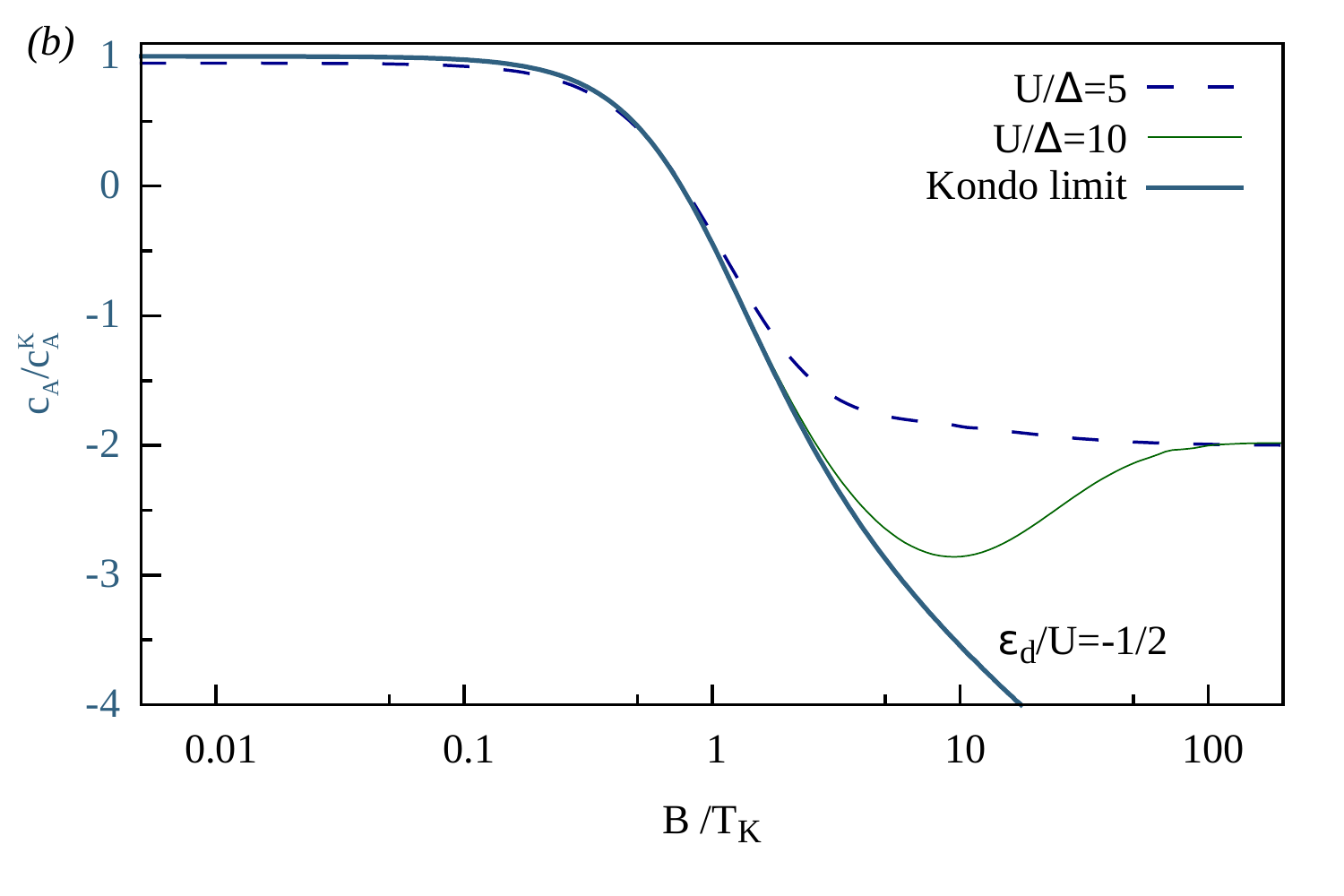}
\includegraphics[width=0.96\columnwidth]{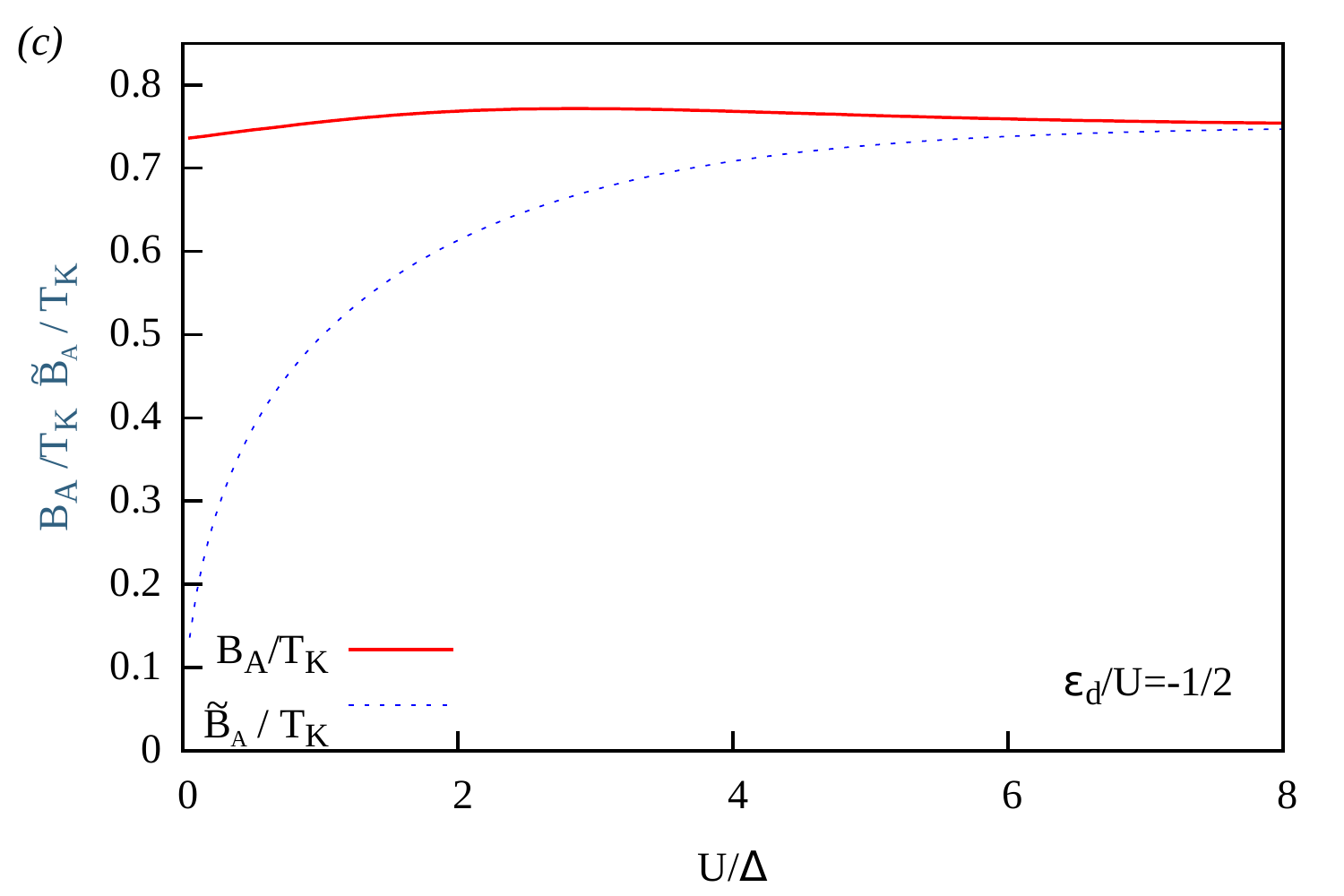}
\caption{\label{fig-kondocA}Low-energy properties of the spectral
  function at particle-hole symmetry.  (a) The normalized height
  coefficient $\tcA/\tcA^K$ and (b) curvature coefficient $\cA/\cA^K$
  of the local spectral function at particle-hole symmetry, plotted as
  functions of magnetic field [in units of $\Tk$, as defined in
  Eq.~\eqref{eq:tksus}], for three values of the interaction parameter
  $U/\Delta$, including the Kondo limit $U/\Delta = \infty$ [given by
  \Eq{eq:tca-ca-Kondo-limit}]. 
  The sign change for $\CA$ signals the splitting of the Kondo
  resonance into two resonances due to the breaking of the Kondo
  singlet by the magnetic field.  (c) The characteristic fields
    $B_A$ and $\tBA$ where $\CA$ and $\tCA$ vanish, respectively,
  plotted in units of $\Tk$ as functions of $U/\Delta$. In the Kondo
  limit $U/\Delta \to \infty$ they approach the same limiting
    value, $B_A = \tBA = 0.75073 \Tk$.}
\end{figure}

To study this behavior quantitatively, we specialize the results
of the previous subsection to the case of particle-hole symmetry using
\Eq{eq:particle-hole-symmetry}, obtaining:
\begin{subequations}
\label{subeq:expandaEpsilon-particle-hole}
\begin{align}
\label{eq:Azero}
A_0 & =  2 \cos^2(\pi m_d) \, , \\ 
\label{eq:tildec_A-prediction-particle-hole} 
\tCA & = 3\phi^2_1\cos (2 \pi m_d)+ \tfrac{1}{2}\bphi_2 \sin (2 \pi m_d) \, , 
\\
\label{eq:c_A-prediction-particle-hole}
\CA  & =  
\left(2 \alpha_1^2+ \phi^2_1\right) \cos (2 \pi m_d) 
         + 2 \balpha_2 \sin (2 \pi m_d) \, .  
\end{align} 
\end{subequations}
Figure~\ref{fig-kondocA} shows the $B$-dependence of
$\tcA = \Estar^2 \tCA$ and $\cA = \Estar^2 \CA$ for several values of
$U/\Delta$. (We multiply by the $B$-dependent scale $\Estar^2$ [cf.\
\Eq{eq:cCEstar}], since this better reveals the large-field behavior,
for reasons explained below.)  
    
The main finding of Fig.~\ref{fig-kondocA} is that with increasing
field, both $\cA$ and $\tcA$ decrease and change sign, as expected from
our qualitative discussion.  The sign change for $c_A$ implies
  that our FL approach reproduces the field-induced splitting of the
Kondo peak in the spectral function.  Moreover, we find
[Fig.~\ref{fig-kondocA}(c)] that the scale for the fields $B_A$ and $\tBA$ is \textit{universal},
in the usual sense familiar from many aspects of Kondo physics in the
Anderson model:  the ratios $B_A/\Tk$ and $\tBA/\Tk$ are of order unity
  and depend only weakly on $U/\Delta$, tending to a constant value
in the Kondo limit $U/\Delta \to \infty$.  Their
limiting value, namely $B_A
  = \tBA = 0.75073 \, \Tk$,
agrees with previous numerical estimates~\cite{costi2000,wright2011}.

Another observation from  Fig.~\ref{fig-kondocA}
is that $\cA$ and $\tcA$ show a very similar field dependence 
for $B/\Tk \lesssim 1$. In the large-field regime their
field dependence differs somewhat for weak interactions,
$U/\Delta \lesssim 5$, but becomes increasingly similar 
with increasing $U/\Delta$. 
  To understand their behavior in the \textit{Kondo limit}
$U/\Delta \to \infty$, we first consider that of 
$\tCA$ and $\CA$.  \Eqs{subeq:expandaEpsilon-particle-hole} 
and \eqref{relations-kondo} show that they are
equal in this limit, given by
\begin{align}
  \label{eq:CA-tCA-Kondo-limit}
\tCA =   \CA = 
  3 \alpha^2_1\cos (2 \pi m_d) 
        +  2 \balpha_2 \sin (2 \pi m_d) \, ,
\end{align}
with zero-field values (indicated by a superscript $K$ for
``fully-developed Kondo effect'') of 
\begin{subequations} 
  \label{subeq:CA-tCA-Kondo-limit}
\begin{align}  
  \label{eq:CA-tCA-Kondo-limit-B=0}
  \tCA^K & = \CA^K = \frac{ 3 \pi^2}{16} \frac{1}{\Tk^2} \quad 
  (B=0) \, ,  
\end{align} 
and asymptotic large-field behavior [obtained 
via \eqref{eq:large-B}] 
\begin{align}  
  \label{eq:CA-tCA-Kondo-limit-largeB}
\tCA =  \CA =   -
\frac{  \pi^2}{B^2 (\ln \beta_r)^4} \ln \beta_r  
\quad   (B\gg \Tk) \, .  
\end{align} 
\end{subequations}
This confirms that $\tCA$ and $\CA$ both become negative at large
  fields. Their magnitude changes in scale from $\sim 1/\Tk^2$ for
small fields to becoming negligibly small,
$\sim 1/[B^2 (\ln \beta_r)^3]$, for large fields.  We now also see why
it is useful to study the $C$ coefficients in the normalized form
$c = \Estar^2 C$ of \Eq{eq:cCEstar}, as done in
Fig.~\ref{fig-kondocA}: $\Estar^2$ increases with $B$ and in the
large-field limit [see \eqref{eq:EstarKondolimit}] compensates the
small prefactor in \Eq{eq:CA-tCA-Kondo-limit-largeB}. Correspondingly
normalized, \Eqs{eq:CA-tCA-Kondo-limit} and
\eqref{subeq:CA-tCA-Kondo-limit} yield
\begin{align}
 \label{eq:tca-ca-Kondo-limit}
\frac{\tcA}{\tcA^K} = \frac{\cA}{\cA^K} 
& =  \cos (2 \pi m_d) +  
 \frac{\partial_B \chi_s}{3 \pi \chi_s^2} \sin (2 \pi m_d) \, , 
\end{align}
with zero-field values and large-field behavior given by 
\begin{subequations}
\begin{align}
\tcA^K  & = \cA^K = \frac{3\pi^2}{16} \, , & (B=0)   \, , \\
\label{eq:tcA-cA-normalized-large-field}
\frac{\tcA}{\tcA^K} & = \frac{\cA}{\cA^K} 
 =
- \tfrac{1}{3} \ln \beta_r  
& (B\gg \Tk) \, .   
\end{align}
\end{subequations}
The $-\ln \beta_r$ term in \Eq{eq:tcA-cA-normalized-large-field}
explains the behavior of the Kondo limit curves (thick solid)
in Figs.~\ref{fig-kondocA}(a) and \ref{fig-kondocA}(b).

As a consistency check, we note that inserting the Kondo-limit
  coefficients $\protect \tCA^K$ and $\protect \CA^K$ of
  \Eq{eq:CA-tCA-Kondo-limit-B=0} into \Eq{eq:expandAepsilon-general}
  for $A(\varepsilon)$ yields the low-energy expansion of the spectral
  function of the spin-$\frac{1}{2}$ Kondo model at $B=0$. Indeed, the
  result so obtained,
\begin{eqnarray}
A^K(\ve) = 2 - \frac{3 \pi^2}{16} \frac{\left[\ve^2 + \frac{1}{3} (\pi T)^2
+ \frac{1}{4}(eV)^2\right]}{\Tk^2} \, , 
\end{eqnarray}
is consistent with previous studies of the Kondo model for $V =0 $
\cite{nozieres1974,affleck1993,glazman2005,Hanl2014} [see for example
Eq.~(4) of Ref.~\cite{Hanl2014}, where the coefficients of this
expansion, called $c_\ve$ and $c'_T$ there, were checked
numerically using NRG].
 
For completeness, we mention that the  opposite limit of 
weak interactions yields, for $\ed = U = 0$: 
\begin{equation}\label{eq:U=0-tCA-CA}
\tcA = 0, \qquad \cA =  \frac{\pi^2}{8} 
\frac{\Delta^2 - 3 B^2/4}{\Delta^2+B^2/4} \, . 
\end{equation}

\subsection{Conductance at particle-hole symmetry}
\label{sec:G-particle-hole}

\begin{figure}[tb]
\includegraphics[width=\columnwidth]{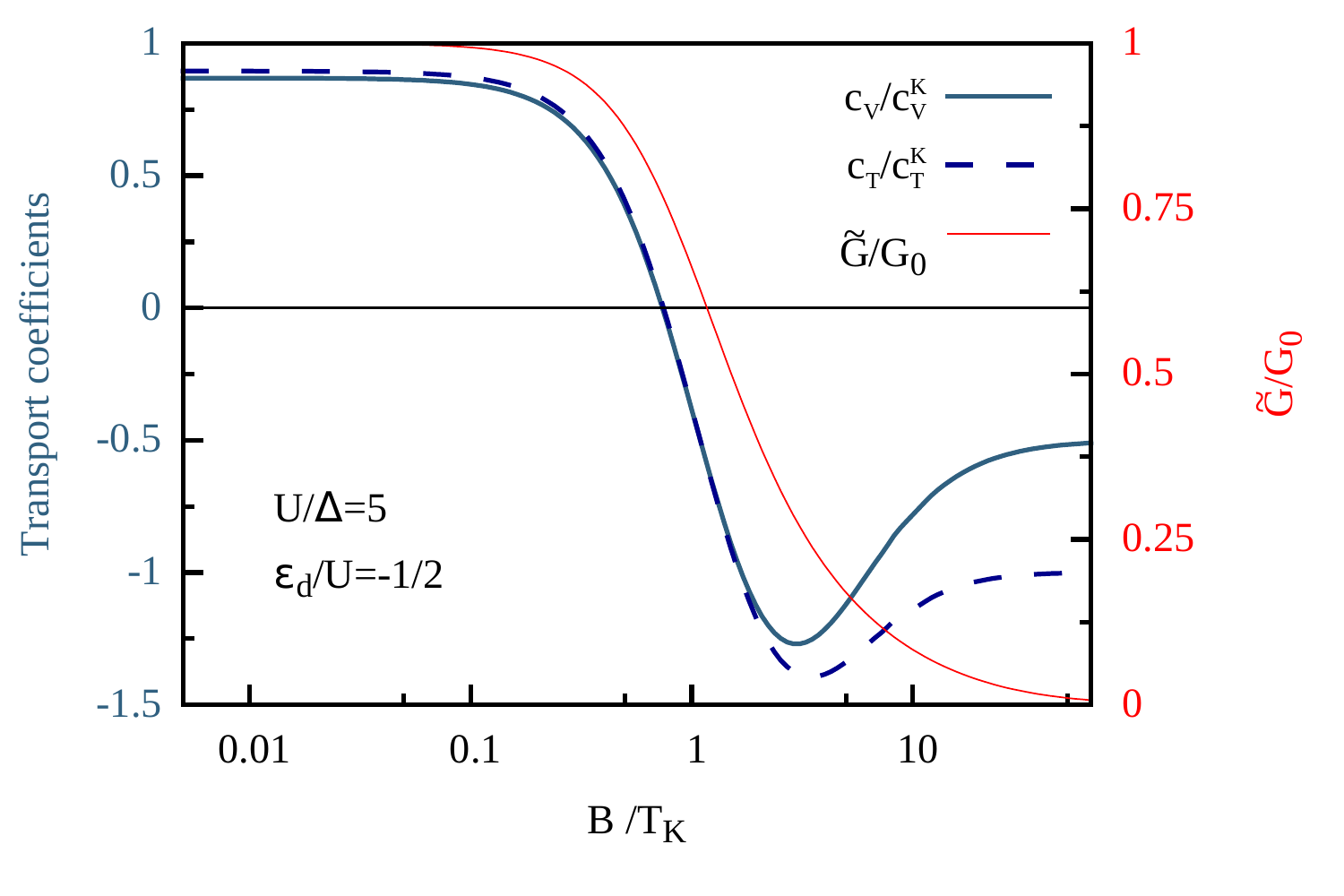}
\caption{\label{fig-cv-cT}FL transport properties at particle-hole
  symmetry and $U/\Delta = 5$, plotted as functions of $B/\Tk$.  Left
  axis: Normalized FL transport coefficients $c_V/c_V^K$ (thick solid
  line) and $c_T/c_T^K$ (thick dashed line).  The fields
    $B_T$ and $B_V$ where $c_T$ and $c_V$ change sign
    are essentially equal and of order $T_K$.
Right axis: Normalized zero-temperature
  linear conductance $\tilde G/G_0 = \cos^2(\pi \md)$ [from
  Eq.~\eqref{eq:Azero}] (thin solid line).}
\end{figure} 

\begin{figure}[tb]
\includegraphics[width=0.9\columnwidth]{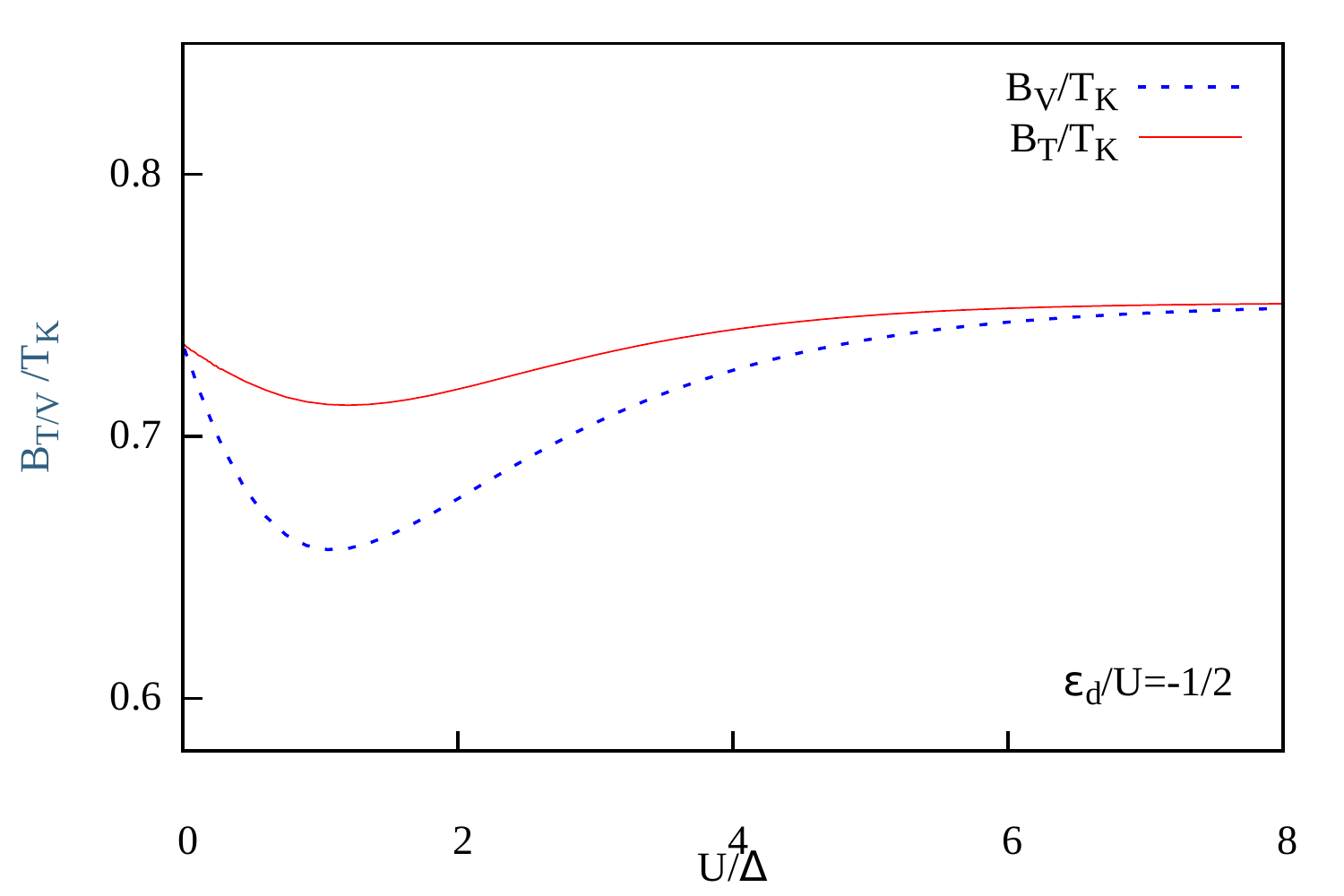}
\caption{\label{fig-vanishing}Interaction dependence of splitting
  field properties at particle-hole symmetry, plotted as functions of
  $U/\Delta$. Splitting fields $B_T$ and  $B_V$ at which $c_T$ and $c_V$ vanish 
  (left axis, thick line), shown in units of $\Tk$.   In
  the absence of interaction $U=0$, the Kondo temperature extracted
  from Eq.~\eqref{eq:tksus} is $T_K = \pi \Delta/2$, hence
  $B_V=B_T = \frac{4}{\pi \sqrt{3}}~T_K\simeq 0.735~T_K$. In the Kondo limit $U/\Delta \to \infty$, both curves approach the limiting value $B_V = B_T = 0.75073~T_K$. 
  The data also agrees with NRG to within the NRG's error
  bars (see Fig.~\ref{fig:B0Fig3}
  in Appendix~\ref{app:NRG}).
}
\end{figure}

We now turn our attention to transport properties, and again begin
with a qualitative discussion.  The behavior of the local spectral
function discussed in the preceding subsection fully determines, via
the Meir-Wingreen formula \eqref{eq:Meir-Wingreen}, that of the
non-linear differential conductance.  At zero field $G(V,T)$, studied
as function of $V$, exhibits a peak around zero bias which weakens
with temperature, and which splits with increasing field.  

 The details of these changes
are quantified by \Eq{eq:cT-cV-cA-ctildeA}: 
as $\tcA$ and $\cA$ decrease with increasing
field and eventually turn negative, the same will
happen for $\cT$ and $\cV$.
  We will denote the ``splitting fields'' at which
$\cT$ or $\cV$ equal zero by $B_T$ or $B_V$, respectively.  Since in
\Eq{eq:cT-cV-cA-ctildeA} the relative weight of $\tcA$ to $\cA$ is
three times larger in $\cV$ than in $\cT$, the behavior of these two coefficients, and that of the
corresponding splitting fields $B_V$ and $B_T$, can thus differ
quantitatively. 

In the noninteracting limit, where $\tcA = 0$, $\cT$, $\cV$ and $\cA$
are proportional to each other for all fields,
$\cV = \frac{3}{4 \pi^2} \cT = \frac{1}{4}\cA$, implying splitting
fields of $B_T=B_V = (2/\sqrt 3)\Delta$ [see \Eq{eq:U=0-tCA-CA}].  At
this field value, the magnetization equals $\frac{1}{6}$ and the
zero-temperature linear conductance is $\tG = \frac{3}{2}e^2/h$, i.e.\
$\frac{3}{4}$ of the unitary value $G_0 = 2e^2/h$.

With increasing $U/\Delta$, the behavior of $\cT$ and $\cV$ 
 are strikingly
similar for small fields and begin to differ only for fields
well above $\Tk$.  This is already evident in
Fig.~\ref{fig-cv-cT}, which shows the $B$ dependence of $\tG$, $\cT$
and $\cV$ for $U/\Delta = 5$. With increasing field, $\tG$ is smoothly
suppressed on a field scale set by $\Tk$, while $\cT$ and $\cV$ both
decrease and change sign, at essentially the same scales: $B_T$ and
$B_V$ are both of order $\Tk$, a property that they directly
inherit from $B_A$ and $\tBA$.
However, the large-field values reached by $c_T$ and $c_V$ for
$B \gg \Tk$ are different. For 
$U/\Delta$ not too large $(\lesssim 5$, as in Fig.~\ref{fig-cv-cT}),
they correspond to the empty-orbital asymptotic forms 
found in Ref.~\onlinecite{mora2015},
\begin{align} c^{\rm eo}_T = -\frac{\pi^4}{16},
\qquad c^{\rm eo}_V = - \frac{3 \pi^2}{64}.
\label{coefficients_in_empty_orbital_regime} 
\end{align} 

A systematic study of the splitting fields $B_T$ and $B_V$ as
  functions of $U/\Delta$ yields the results shown in
  Fig.~\ref{fig-vanishing}. They show qualitatively similar behavior,
  remaining of order $\Tk$ for all values of $U/\Delta$. This implies
  that $B_T$ and $B_V$ are \textit{universal} in the same sense as
  $B_A$.
  
In the \textit{Kondo limit}, \Eqs{eq:cT-cV-cA-ctildeA}, 
\eqref{eq:CA-tCA-Kondo-limit} and \eqref{subeq:CA-tCA-Kondo-limit} yield
\begin{equation}
\label{subeq:CT-CV-Kondo-limit}
\frac{\cT}{\cT^K} 
= \frac{\cV }{\cV^K}  
 = \cos (2 \pi m_d) +
 \frac{\partial_B \chi_s}{3 \pi \chi_s^2} \sin (2 \pi m_d) ,
\end{equation}
 with zero-field values and large-field behavior given by
\begin{subequations}
\begin{align}
\cT^K  & = \frac{\pi^4}{16} ,  \qquad  
\cV^K   = \frac{3 \pi^2}{32} &   (B=0)  , 
\label{eq:cv_ct_Kondo} 
\\
  \label{eq:CT-CB-Kondo-limit-largeB}
\frac{c_T}{\cT^K} & = \frac{c_V}{\cV^K} =
-  \tfrac{1}{3} \ln \beta_r
& (B\gg \Tk) . 
\end{align}
\end{subequations}

\begin{figure}[tb]
\includegraphics[width=\columnwidth]{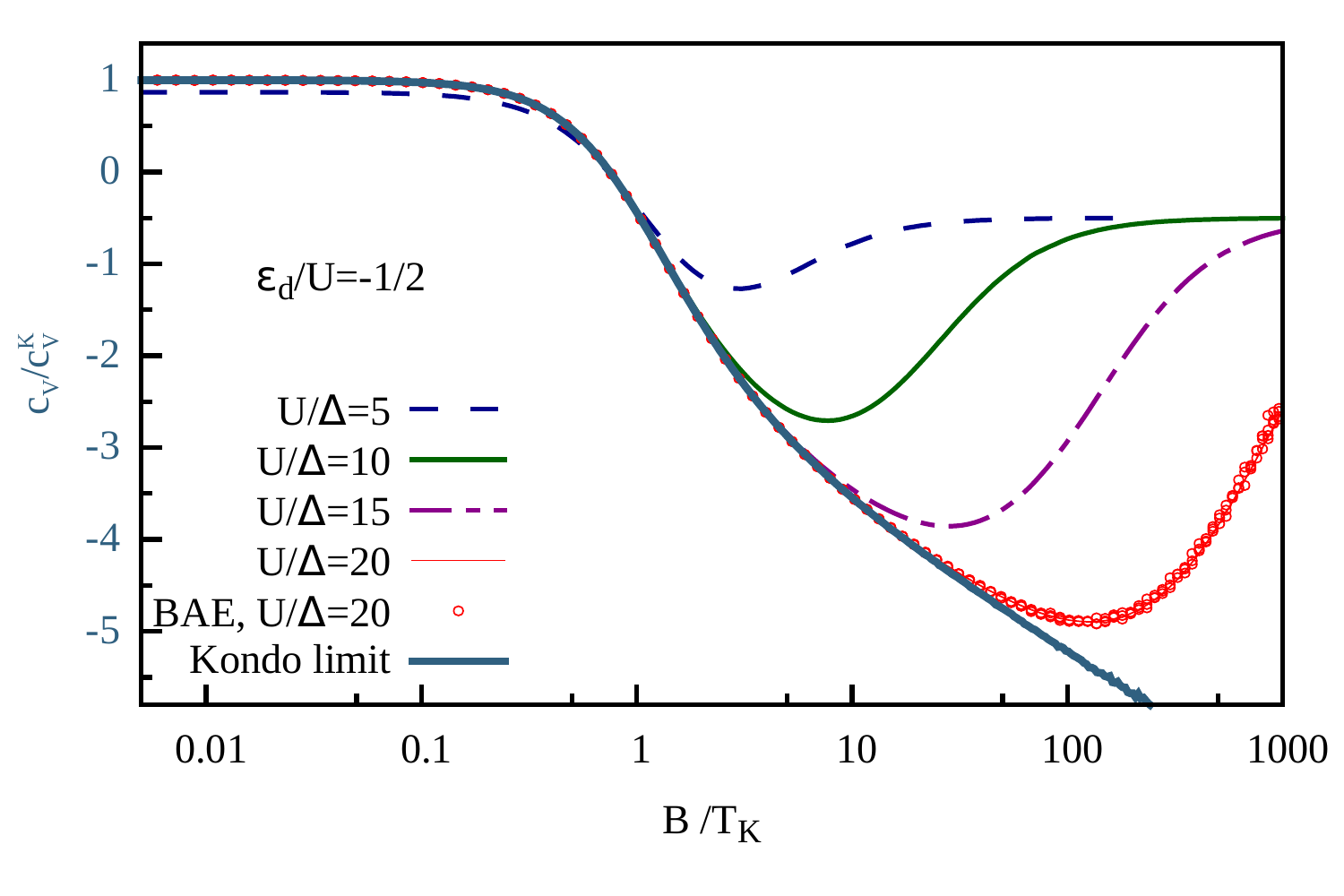}
\caption{\label{fig-kondocv}Evolution of $\cV$ during
  the crossover to the Kondo limit. 
 The normalized FL transport
  coefficient $c_V/c_V^K$ is plotted as a function of $B/\Tk$ for
  several values of $U/\Delta$, 
  including the Kondo limit (thick solid line).  
   The direct
  integration of the coupled Bethe Ansatz equations, Eq.~(S3a)
  and~(S4b)~\cite{Note1}, performed here for $U/\Delta=20$ (BAE, open dots),
  is in very good agreement with the corresponding Wiener-Hopf
  solution (dashed line). All curves with large
$U/\Delta$ ($\gtrsim 10$) initially collapse onto a universal scaling
curve as function of increasing  $B/\Tk$, but for large $B/\Tk$ they eventually bend upward towards
zero at a field scale that increases with $U/\Delta$
and tends to infinity in the Kondo limit.}
\end{figure}

\begin{figure}[tb]
\includegraphics[width=\columnwidth]{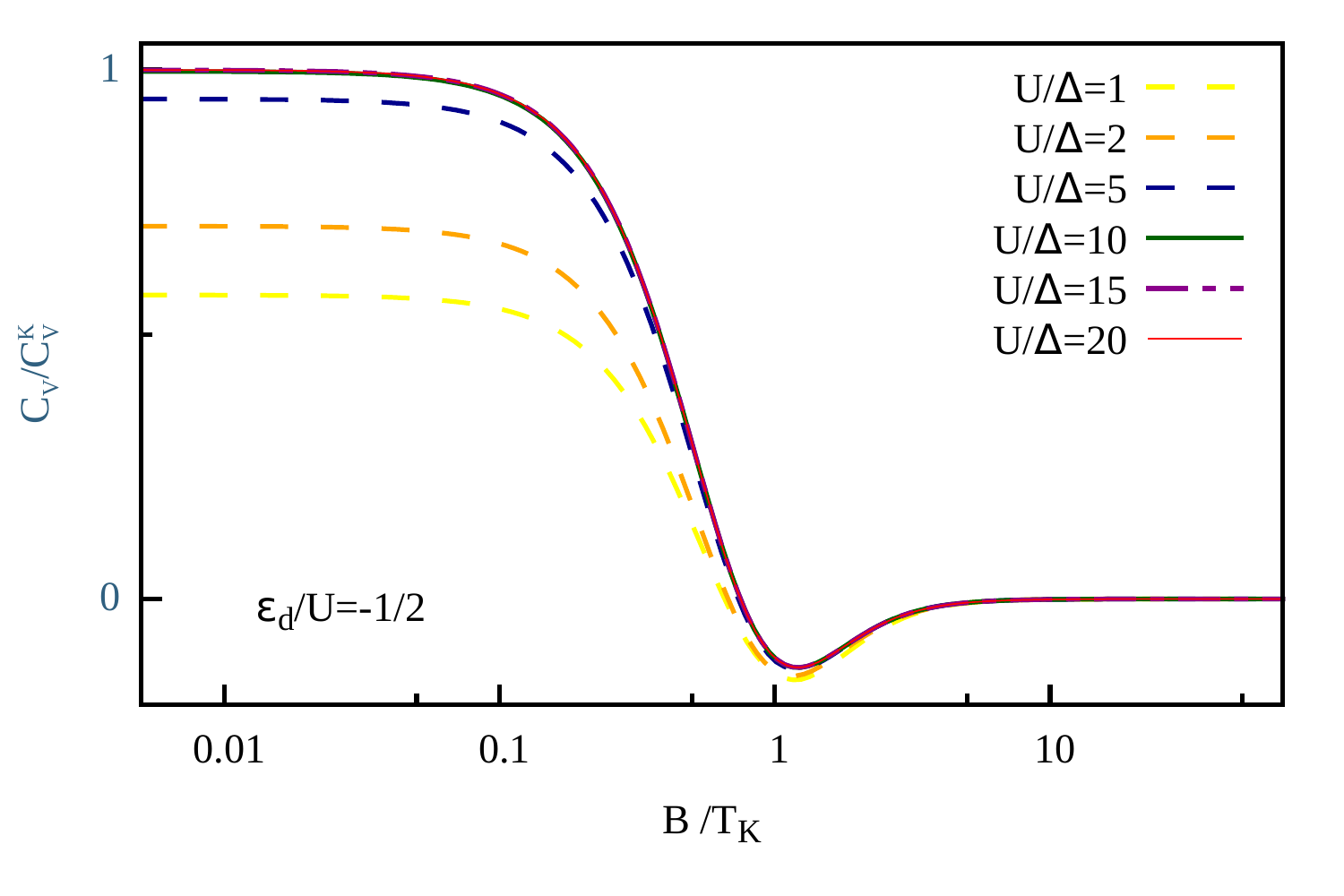}
\caption{\label{fig-BIGCV} The coefficient $\CV = \cV/\Estar^2$,
      plotted as a function of $B/T_{\rm K}$ for different interaction
      strengths $U/\Delta$, in units of
      $\CV^{\rm K}=\cV^{\rm K}/\Tk^2$, with $\cV^{\rm K}$ defined in
      Eq.~\eqref{eq:cv_ct_Kondo}.  In contrast to $\cV$ from
      Fig.~\ref{fig-kondocv}(a), $C_V$ is strongly suppressed in the
      regime $B>\Tk$ (even for $U/\Delta\gg 1$), because for large
      fields the spin susceptibility becomes very small and hence
      $\Estar$ very large [cf.\
      Eq.~\eqref{eq:EstarKondolimit}]. Inset: the normalization factor
      $\CV^{\rm K}$, plotted as function of $U/\Delta$ in units of
      $1/\Delta^2$ (black points indicate the $U/\Delta$ values from
      the main plot). $\CV^{\rm K}$ shows an exponential increase
      with $U/\Tk$, caused by an exponential decrease in $\Tk$.
      The growth in $\CV^{\rm K}$ is 
      counteracted by the fact that the voltage window in which our FL
      analysis applies decreases exponentially, since the FL expansion
      \eqref{eq:c's-B-dependent} of the conductance requires
      $V\ll \Tk$.}
\end{figure}\label{footnote:hewson2005} 

To conclude this section, we remark that it is instructive to compare
  the predictions of our FL theory  with those of Hewson, 
  Bauer and Oguri \cite{hewson2005}, who computed $B_V$ for the case
  of particle-hole symmetry using renormalized perturbation theory
  (RPT) [see the discussion after their Eqs.~(19) and (27)]. For
  example, for $U/\Delta = 4 \pi$ they find
  $\tfrac{1}{2} B_V \simeq 0.584 \Tk$ [after their Eq.~(27)].  This is
  comparable in magnitude, but not equal, to our result
  $B_V \simeq  0.7506 \, \Tk$ [see our
  Fig.~\ref{fig-vanishing}] for the same value of $U/\Delta$.  
However, we note that in Ref.~\cite{hewson2005} the
coefficients playing the roles of our $\alpha_{2 \sigma}$ and
$\phi_{2 \sigma}$ were computed perturbatively in terms of the
renormalized parameters of RPT (the \textit{same} three parameters are
also used at zero magnetic field~\cite{hewson2005b}), and are
therefore approximate.  As noted in the concluding section of
Ref.~\cite{mora2015}, it is not clear whether this RPT approach
contains enough parameters to accurately evaluate $B_V$.

In an attempt to track down the difference,  we have expressed
  our FL parameters in terms of the RPT parameters needed in general
  to characterize the local impurity Green's function (see
    Appendix~\ref{app:RPT}). This can be done by simply expanding the
  RPT spectral function to second order in $\ve$, $T$ and $eV$ and
  equating the result to our Eqs.~\eqref{eq:expandAepsilon-general}
  and \eqref{subeq:Acoefficients}. The resulting equations
  \eqref{eq:FL-RPT-general-dictionary} provide a RPT-FL dictionary
  that relates the RPT parameters to our FL parameters. Since the
  latter are computable exactly via the Bethe Ansatz, this dictionary
  provides a number of exact constraints on the RPT parameters. We
  were not able to ascertain that the expressions provided in
  Ref.~\cite{hewson2005} for their RPT parameters satisfy these
  constraints. We suspect that at finite magnetic field 
  or out of particle-hole symmetry, the second-order RPT 
(perturbative in the renormalized interaction $U$) becomes approximate 
  for the calculation 
  of the coefficients $\alpha_2$ and $\phi_2$ \cite{Hewson2016}.
 However, we would like to suggest a converse strategy:
  one could set up a RPT whose input parameters are computed exactly
  by Bethe Ansatz via the RPT-FL dictionary in
  Appendix~\ref{app:RPT}. Doing so would be an interesting goal for
  future work, since RPT offers the welcome prospect of smoothly
  linking the exact FL description of the impurity's low-energy
  behavior to a description, albeit approximate, that is also useful
  at higher energies.

  Very recently, Oguri and Hewson have taken a decisive step
    forward which completes the RPT program and puts it on a fully
    rigorous footing \cite{Oguri2017,*Oguri2017a,*Oguri2017b}: they
    used Ward identities together with the analytic and antisymmetry
    properties of the vertex function of the Anderson impurity model
    to fully determine all parameters needed in RPT. In doing so, they
    also presented a microscopic derivation of the FL relations
    arising from a low-energy expansion of the self-energy and the
    vertex. Their results are in full agreement with the FL theory
    presented in section II of this work.  Indeed, we show in
      appendix~\ref{app:oh} that our analytical expressions for $C_T$
      and $C_V$ coincide, for general values of $\varepsilon_d$, with
      those obtained by them. For the case of particle-hole symmetry,
      they combined their FL analysis with NRG computations of the FL
      parameters. Their  prediction for the splitting field for the zero-bias
      conductance peak is $B_V/2 = h_V \simeq 0.4 \Tk$ for
    $U/\Delta = 4 \pi$, implying $B_V/\Tk \simeq 0.8$. This is in
    fairly good agreement with our prediction at $U/\Delta = 4 \pi$,
    namely $B_V/\Tk = 0.7506$.  The difference is likely due
      to different methods for numerically determining the FL
      parameters -- NRG in their work, Bethe Ansatz in ours.
 

As another consistency check for our FL analysis, we
have used NRG to compute the equilibrium spectral 
function $A(\epsilon)$ and  extracted
$C_T$ and $C_V$ from its leading 
dependence on $T$ and $\epsilon$. The results of this
analysis, presented in Appendix~\ref{app:NRG}, 
are consistent with the FL results 
for $C_T$ and $C_V$ of Fig.~\ref{fig-cv-cT},
and $B_T$ and $B_V$ of Fig.~\ref{fig-vanishing}.

\begin{figure}[h!]  
\begin{center}
\includegraphics[width=.48\textwidth]{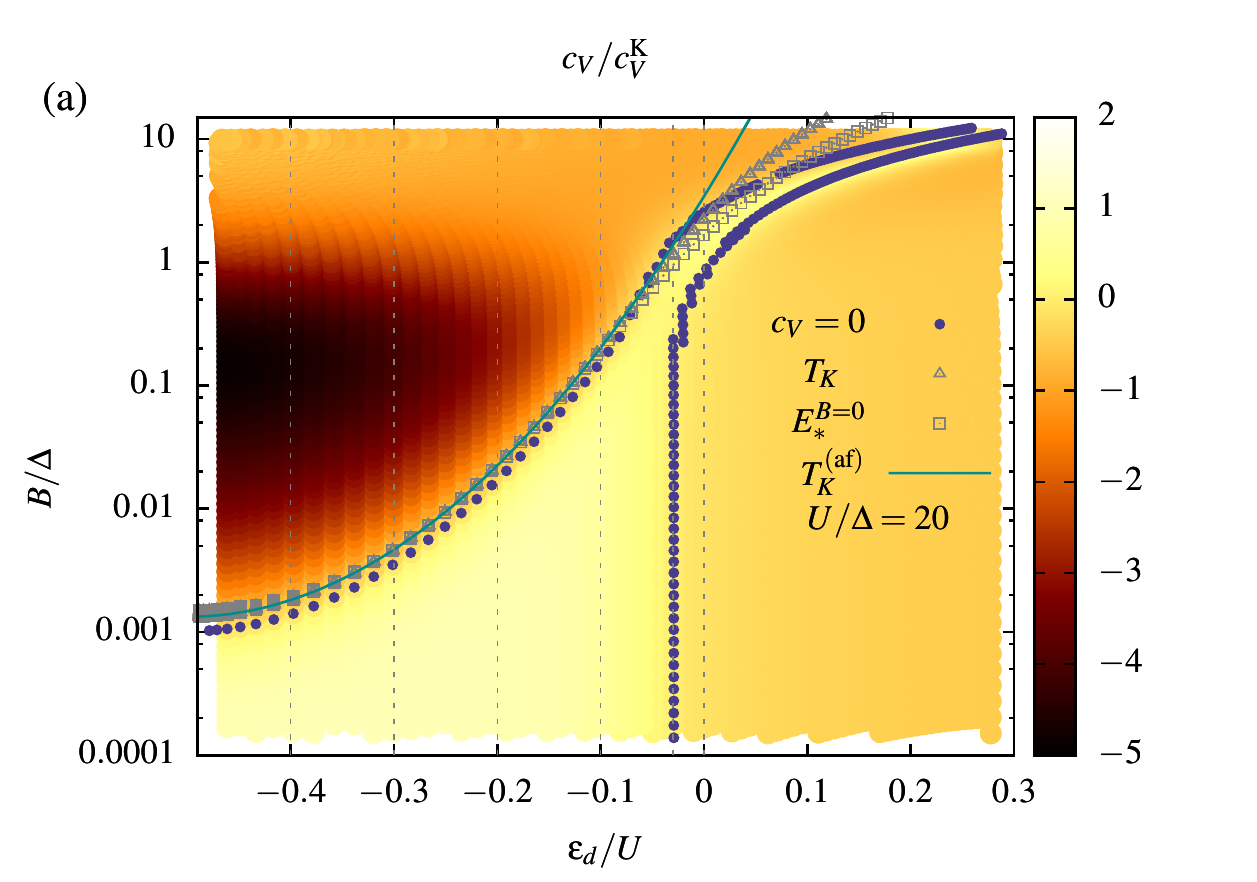}
\includegraphics[width=.47 \textwidth]{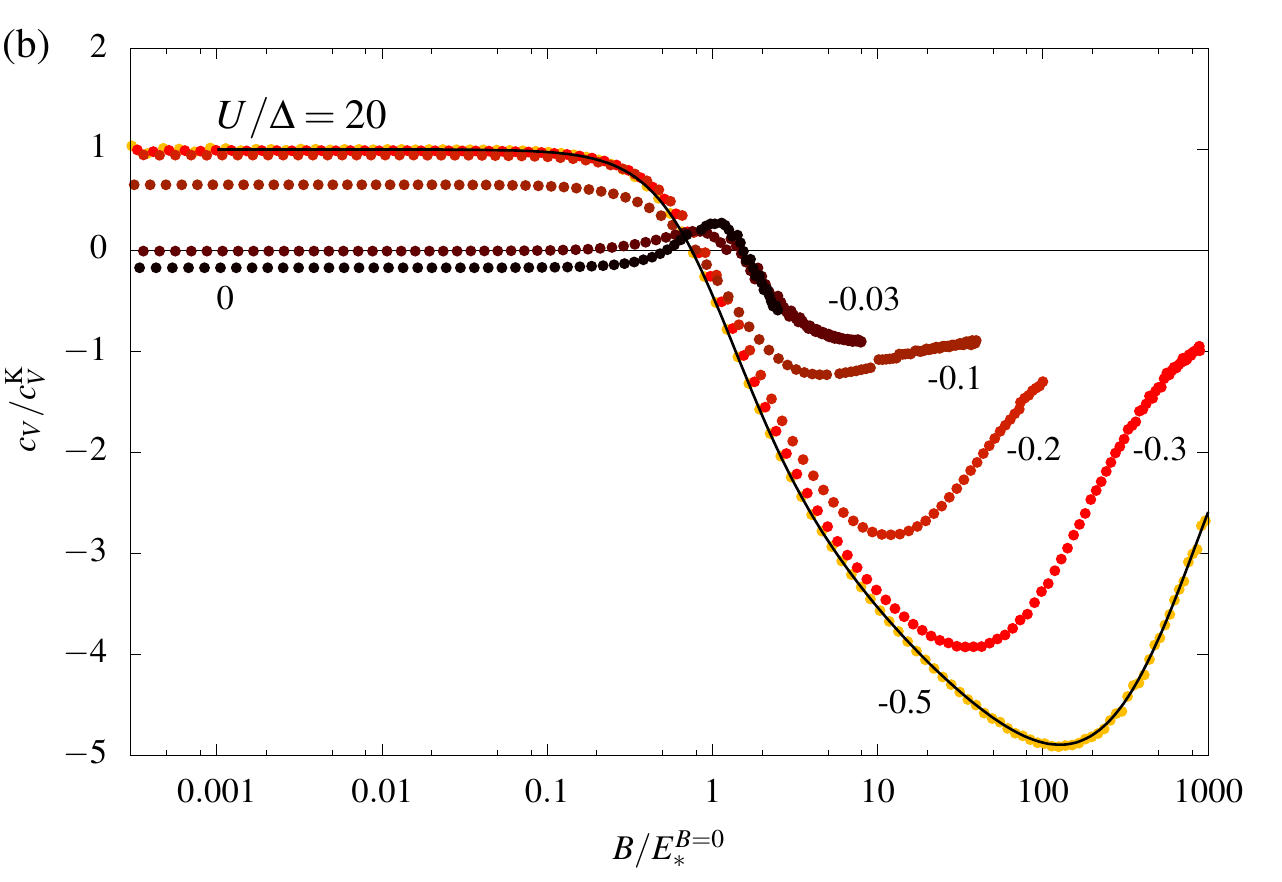}
\caption{(a) Transition from the local-moment regime to
  the empty-orbital regime for the transport coefficient $c_V/c_V^K$,
  shown using a color scale, as a function of the magnetic field $B$
  and the level energy $\varepsilon_d$, at $U/\Delta=20$, a convenient
  value to highlight features related to Kondo physics. The solid
    line shows the prediction for the Kondo scale of the analytic
    formula \eqref{eq:estar} for $\Tkaf$, the grey
    triangles the numerical evaluation of $\Tk$ as defined in
    Eq.~\eqref{eq:tksus} and the grey squares the numerical evaluation of $E_*$ in Eq. \eqref{eq:define-Estar} for $B=0$. All these quantities show a nice agreement as long as
  $\varepsilon_d<0$. 
The  blue points signal when $c_V=0$ and changes
  sign.  
The light-colored regions correspond to positive values of
  $c_V$.  (b) Same quantity $c_V/c_V^K$ as in (a), now shown
  along the cuts marked in (a) by grey dashed lines, and plotted
  as a function of $B/E_*^{B=0}$ on a logarithmic scale. The numbers
  above the data points give the corresponding values of
    $\varepsilon_d/U$ (increasing as colors turn from light to dark).
  The solid line corresponds to the analytical result for $c_V$ at particle-hole symmetry   derived from the Wiener-Hopf solution \cite{Note1} (see also Fig. \ref{fig-kondocv}a).  Throughout the local-moment regime in which Kondo correlations occur ($\varepsilon_d \lesssim - \Delta$), $c_V$ changes 
    sign around fields of order $\Tk$. For
    $\varepsilon_d\gtrsim0$, $c_V$ develops a double sign change with
    a peak in between, which reflects a field-induced resonance
    between the empty- and singly-occupied dot states (see
    Fig.~\ref{fig:cvcuts}).
} \label{fig:cvdens} \vspace{-5mm}
\end{center}
\end{figure} 

\subsection{$c_V$ away from particle-hole symmetry}\label{sec:oph}

Finally, let us examine the behavior of the transport coefficient
$c_V$ away from particle-hole symmetry. We consider only
$\varepsilon_d/U>-\frac{1}{2}$ (from which the opposite case follows
by particle-hole symmetry).  The quantum dot is in a strongly
correlated Kondo singlet state as long as the dot is in the
local-moment regime ($-U/2 \le \varepsilon_d \lesssim - \Delta$). As
$\varepsilon_d$ crosses over through the mixed-valence regime
($|\ed| < \Delta$) into the empty-orbital regime
($\ed \gtrsim \Delta$), Kondo correlations die out completely.  In the
previous section, we showed that $c_V$ changes sign at particle-hole
symmetry for splitting fields $B_V$ of the order of
$\Tk$ [see
Fig.~\ref{fig-vanishing}(b)]. Our aim here is to study the evolution
of $B_V$ as $\varepsilon_d/U$ is tuned through the transition from the
local-moment regime to the empty-orbital regime. The numerical results
reported below were obtained by numerically solving the Bethe-Ansatz
equations for the Anderson model
\cite{wiegmann1980,*kawakami1982ground,*wiegmann1983,*tsvelick1983,*tsvelick1983a}
in the form reported in the Supplemental Material~\cite{Note1}.

Fig.~\ref{fig:cvdens}(a) shows a color-scale plot of $c_V$ for a
large, fixed interaction of $U/\Delta = 20$, plotted as function of
field $B$ and level energy $\varepsilon_d$.  Fig.~\ref{fig:cvdens}(b)
shows the same data as function of $B$ along several fixed values of
$\ed$. We find that throughout the local-moment regime, an increasing
field yields a sign change for $c_V$ as
function of $B$ around field values that distinctly follow the $\ed$
dependence of the Kondo temperature $\Tk$ of \Eq{eq:tksus} (grey
triangles). The latter is well approximated by the analytic formula
$\Tkaf$ of \Eq{eq:estar} (black solid line) and coincides with the FL
scale $E_*$ of Eq. \eqref{eq:define-Estar} at $B=0$ (grey squares),
with deviations only in the empty orbital regime
($\ed \gtrsim \Delta$).  

\begin{figure}
\begin{center}
\includegraphics[width=.47\textwidth]{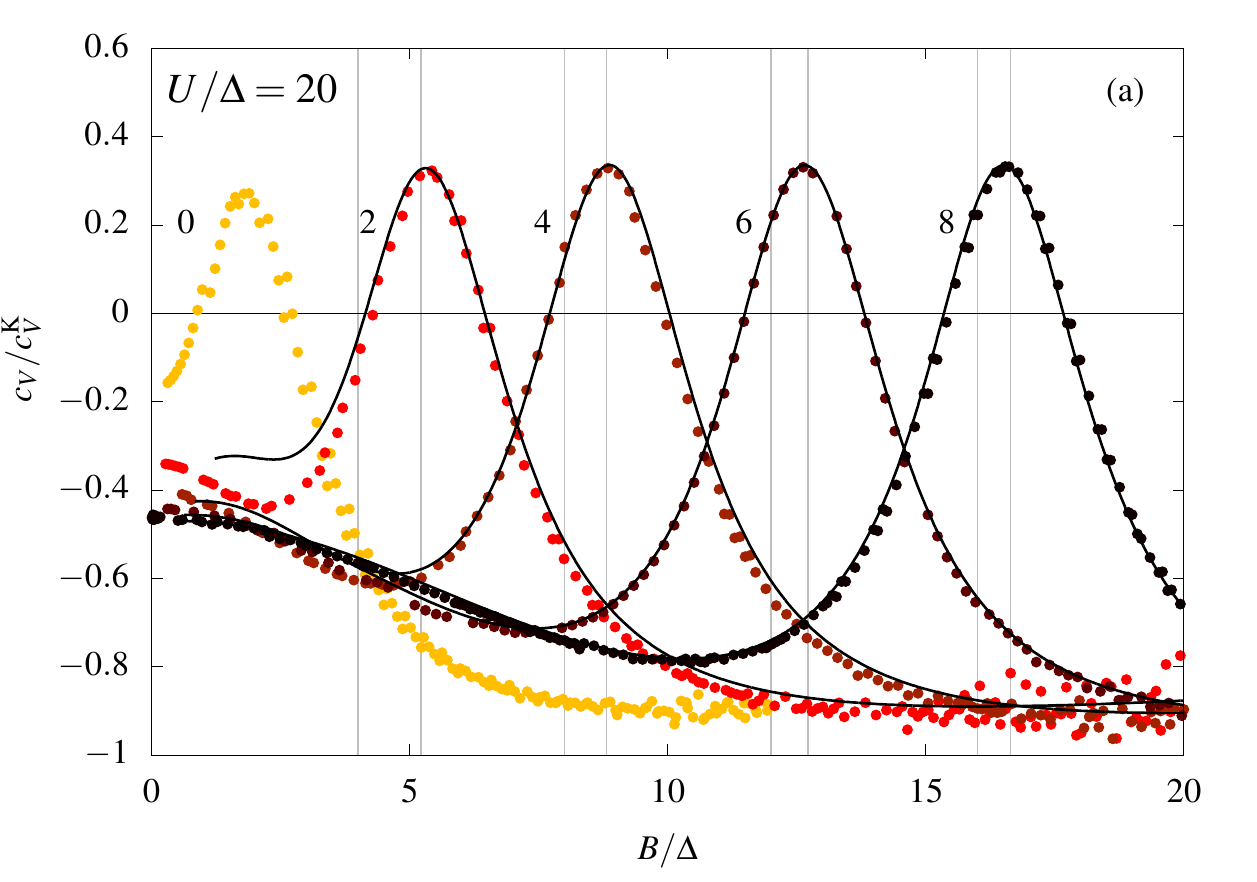}
\includegraphics[width=.47\textwidth]{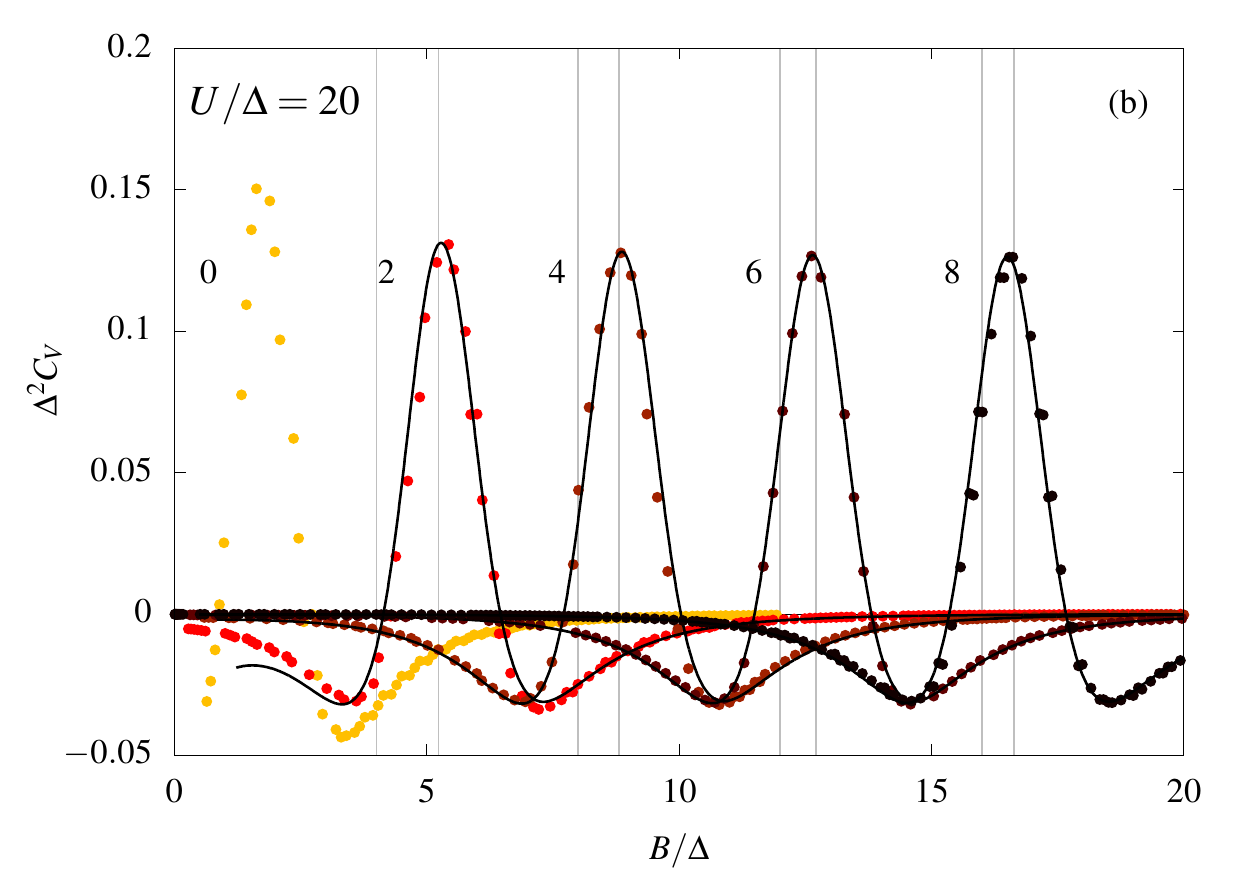}
\caption{ Magnetic-field dependence of (a) $c_V$ in units of
    $\cV^{\rm K}$ and (b) $\CV=\cV/\Estar^2$ in units of $1/\Delta$,
    in the mixed-valence and empty-orbital regimes. Both are plotted
  as functions of $B/\Delta$ on a linear scale, for several values of
  $\varepsilon_d/\Delta$ (given by numbers above the data points,
  increasing as colors turn from light to dark). Black solid curves
  display analytical predictions derived in perturbation theory,
  showing good agreement with the numerical results (symbols). The
  peaks in $\cV$ and $\CV$ reflect the field-induced resonance between
  the empty and singly-occupied dot states. Vertical solid lines
  indicate the predicted values of the resonance field,
  $B=2\tilde\varepsilon_d$, where the renormalized level position
  $\ted$ is given by \Eq{eq:edtilde} (and $\alpha \simeq 1.62$
  therein); for comparison, vertical dashed lines indicate the bare
  values, $B=2\varepsilon_d$. Note that the non-trivial $B$-dependence
  exhibited by $\CV$ in (b) is as pronounced as that of $\cV$ in
  (a). The reason is that near the resonance field
  $B \simeq 2\tilde \varepsilon_d$, both the spin and charge
  susceptibilities are large, ensuring that $\Estar$ remains small.}
\label{fig:cvcuts} \vspace{-5mm}
\end{center}
\end{figure}

The behavior of $c_V$ is strongly modified as soon as the renormalized
level increases past the Fermi surface
($\varepsilon_d \gtrsim \Delta$) and the charge on the dot changes
from 1 to 0, so that Kondo correlations are completely absent. For low
magnetic fields, $c_V$ is negative and with increasing field evolves
through a double sign change with a positive-valued peak in between,
see Figs. \ref{fig:cvdens} and \ref{fig:cvcuts}.  This behavior can be
understood as follows.  At zero magnetic magnetic field, the dot is
empty and in a cotunneling regime \cite{grabert2013}, so that its
  conductance increases when the bias increases from zero. This
  explains why $c_V$ is found to be negative for small fields in
  Fig.~\ref{fig:cvdens}.  With increasing field, the local level is
  Zeeman split.  When the empty and singly-occupied states come into
  resonance, the conductance develops a well-pronounced zero-bias peak
  with a negative curvature, explaining why $c_V$ goes through a
  positive-valued maximum.  The resonance condition at which this
  happens is that the spin-up Zeeman energy $B/2$ matches
  the renormalized level position $\ted$, which 
  differs from $\ed$ due to virtual processes involving
  doubly-occupied intermediate dot states.
A perturbative calculation following Haldane \cite{haldane1978} yields
\footnote{Eq.~\eqref{eq:edtilde} is derived from the seminal
  expression of the renormalized level provided by Haldane in 
  Ref. \cite{haldane1978},
$\varepsilon^\ast_d=\varepsilon_d+\frac\Delta\pi\ln\frac{W_0}{W}$,
in which $W_0$ is the energy scale at which the orbital energy
renormalization starts, namely $W_0\sim \varepsilon_d+U$, and $W$ the
energy scale in which the renormalization stops, \textit{i.e.}
$W\sim\alpha\varepsilon_d$. Notice that this formula 
for $\varepsilon^\ast_d$ applies within a 
regime different from the mixed valence regime; in the latter,
  $|\varepsilon_d| < \Delta$, hence $\varepsilon_d$ is the lowest
energy scale in the problem and the renormalization stops at
$\Delta\ll U$. For the data sets presented in  Fig.~\ref{fig:cvcuts}(a),  
$\varepsilon_d$ is still comparable 
to (but somewhat larger than) the
hybridization energy $\Delta$ and the charging energy $U$.}
\begin{align}\label{eq:edtilde}
\tilde\varepsilon_d=\varepsilon_d+\frac\Delta\pi\ln\frac{\varepsilon_d+U}{\alpha \varepsilon_d}\,,
\end{align} 
where $\alpha$ is a constant of order one. For the
choice $\alpha \simeq 1.62$ the resonance field values
  $B=2\tilde\varepsilon_d$, indicated by vertical solid lines in
  Fig.~\ref{fig:cvcuts}, indeed match the observed
peak positions for $c_V$ rather well.

For $\varepsilon_d>\Delta$, the full dependence of $c_V$ on the
magnetic field can be well captured analytically by second-order
perturbation theory in the dot-lead hybridization
\cite{filippone2013}, using the spin-down state of the dot as virtual
intermediate state. Appendix \ref{app:largeed} presents
  corresponding results for $\ndsigma$ as function of the bare level
  position $\ed$ and Zeeman field $B$, from which $c_V$ can be
  obtained using the formulas of Sec.~\ref{sec-FL}. Using the
  substitution $\varepsilon_d\rightarrow\tilde\varepsilon_d$ in the
  final results, one obtains the solid curves for $c_V$ shown in
  Fig.~\ref{fig:cvcuts}, which agree nicely with our numerical results
  (symbols). In the limit $\varepsilon_d\gg\Delta$, where the
  spin-down state can be totally neglected, the shape of the $c_V$
  peak can be computed by considering a single non-interacting
  resonant level. The result is 
\begin{align}\label{eq:cvscalingbesser}
\frac{c_V}{c_V^{\rm K}}&=\frac13\frac{\Delta^2 -
3\left(\tfrac{1}{2} B - \ed\right)^2}{\Delta^2 +\left(\tfrac{1}{2}B - \ed\right)^2}\, ,
\end{align}
which is peaked symmetrically around the resonance field $B = 2 \ed$.

In the mixed-valence and empty-orbital regimes, the non-trivial
  $B$-dependence exhibited by $\cV$ is equally well visible in
  $C_V=c_V/E^{*2}$, see Fig.~\ref{fig:cvcuts}(b) (in contrast to the
  Kondo regime, where $\CV$ rapidly approaches zero for
  $B \gtrsim \Tk$, cf.\ Fig.~\ref{fig-BIGCV}). The reason is that in
  the mixed-valence and empty-orbital regimes the FL scale $E_*$ does
  \textit{not} become very large with increasing $B$, because both the
  spin and charge susceptibilities $\chi_s$ and $\chi_c$ are sizable,
  ensuring that $\Estar$ remains small [cf.\
  Eq.~\eqref{eq:define-Estar}].  In fact, both susceptibilities become
  maximal, and $\Estar$ minimal, in the regime near near
  $B \simeq 2 \tilde \varepsilon_d$ where the empty and
  singly-occupied dot states are in resonance. This can be checked
  analytically in the $\varepsilon_d\gg \Delta$ limit, where the
  perturbative approach presented in Appendix \ref{app:largeed} yields
\begin{equation}
E_{*}=\frac \pi{2\Delta}\Big[(\varepsilon_d-B/2)^2+\Delta^2\Big]\,,
\end{equation}
which is \textit{minimal} at the bare resonance field $B = 2 \ed$. 

The fact that $\CV$ is large in the mixed-valence and empty-orbital regimes
suggests that these regimes would be particularly suitable
for the purposes of benchmarking numerical methods for
solving the nonequilibrium Anderson model against the
exact results obtained by our FL approach. 

\section{Summary and conclusions} \label{sec-conclusion}

We extended the FL framework of Ref.~\onlinecite{mora2015} to the
single-impurity Anderson model at finite magnetic field where low
energy properties can be calculated in the whole phase diagram.  Using
a generalization of the ``floating Kondo resonance'' argument of
Nozi\`eres, we expressed all parameters of the low-energy effective FL
Hamiltonian in terms of the zero-temperature local occupation
functions $\ndsigma$ and their derivatives with respect to level
energy and magnetic field. Our results are in full agreement
  with a recent analysis of Oguri and Hewson
  \cite{Oguri2017,*Oguri2017a,*Oguri2017b}. We evaluated our
expressions for the FL parameters using precise Bethe Ansatz
calculations. Focussing on strong interaction, zero temperature and
particle-hole symmetry where the Kondo singlet forms, we obtained
exact results for the magnetic-field dependence of $\tcA$ and $\cA$,
two parameters that characterize the zero-energy height and curvature
of the equilibrium spectral function, respectively.
We also computed the
  splitting field $B_A$ at which $\cA$ changes sign, signaling the
  onset of a field-induced splitting of the equilibrium Kondo peak,
and find that $B_A$ is of order $\Tk$ throughout the local-moment
regime, as expected.

We next performed exact calculations of the FL transport coefficients
$c_T$ and $ c_V$ at particle-hole symmetry but for arbitrary magnetic
fields. In the local-moment regime we find that  both $c_T$ and $c_V$ change sign at a field of
  order $\Tk$, as expected. Finally, we also calculated the magnetic-field dependence of $c_V$
throughout the crossover from the local-moment through the
mixed-valence into the empty-orbital regime.  Throughout the former,
the behavior of $c_V$ is qualitatively similar to that found at
particle-hole symmetry. However, it changes dramatically upon entering
the empty orbital regime: there $c_V$ is negative at zero magnetic
field, but with increasing field traverses a positive-valued peak at
twice the renormalized level energy, $B \simeq 2 \ted$, arising from a
spin-polarized resonance between the empty and singly-occupied dot
states.

It would be an interesting challenge for experimental studies of
nonequilibrium transport through quantum dots to check our
peak-splitting predictions by detailed measurements of the nonlinear
conductance as function of bias voltage and field. Since the specifics
of our peak-splitting predictions are model dependent, it would be
important to strive for a faithful implementation of the
\textit{single-level} Anderson model, requiring a small dot with a
very large level spacing, and to make the ratio $U/\Delta$ as large as
possible.

To conclude, we have used exact tools to address the question posed in
the title of our paper, 
finding $B_T = B_V = 0.75073 \Tk$ in the Kondo limit. On a
quantitative level, our work establishes exact benchmark results
against which any future numerical work on the nonequilibrium
properties of the Anderson model can be tested.

\section*{acknowledgments}
We thank S. Arlt, M. Kiselev, H. Schoeller, D. Schimmel,
A. Weichselbaum and L. Weidinger for helpful discussions, and
A. Hewson for several very helpful exchanges on renormalized
perturbation theory.  In particular, we would like to thank A. Oguri
and A. Hewson for a very helpful exchange \cite{OguriPrivate2017}, which
lead us to discover the sign error in the computation of the
conductance in Ref.~\cite{Filippone2017}.  This work has been
supported by the Cluster of Excellence ``NanoSystems Initiative
Munich''.

\appendix

\section{Kondo model at large magnetic field}\label{appen:kondo}

In this appendix, we perform a consistency check of the FL theory of
the main text by considering the Kondo model in the large-field limit
$B\gg \Tk$.  
To this end, we derive an effective FL Hamiltonian from the Kondo
Hamiltonian by doing second-order perturbation theory in spin-flip
scattering. This yields explicit expressions, in terms of the
\textit{bare} parameters of the Kondo model, for the FL parameters
$\delta_{0\sigma}$, $\alpha_{j\sigma}$ and $\phi_{j\sigma}$, and
hence, via the FL relations \eqref{fermic}, also for
  $m_d$, $\chi_s$ and $\partial \chi_s/\partial B$.
Satisfyingly, the latter
  expressions turn out to fully agree with corresponding Bethe-Ansatz
  results in the large-field limit.

A standard mapping exists between the Anderson model
  Eq.~\eqref{am} and the Kondo model
  $H_K = - B \, S_z + \sum_{\sigma,k} \varepsilon_{k}
  c^\dagger_{k\sigma} c_{k\sigma} + H_{\rm ex}$
  with the spin-exchange interaction
\begin{equation}
 H_{\rm ex} = J \, {\bf S} \cdot {\bf s},
\end{equation}
where
$\mathbf s=\sum_{kk'\sigma\sigma'}c^\dagger_{k\sigma}\frac{{\bm
    \tau}_{\sigma\sigma'}}2c_{k'\sigma'}$
denotes the local spin of conduction electrons and
${\bm \tau}_{\sigma\sigma'}$ is a vector composed of the Pauli
matrices. The mapping holds at particle-hole symmetry, where
$\nu_0 J = 8 \Delta/(\pi U)$, and for energies well below the charging
energy $U$. The impurity then hosts exactly one electron with spin
$\mathbf S$.

The FL Hamiltonian of Eq.~\eqref{eq:H_FL} can be
derived perturbatively at large magnetic fields $B \gg T_K$. A strong
magnetic field polarizes the impurity and a perturbation expansion with respect
to the impurity in the spin up state $|\! \uparrow \rangle$ can be
formulated. The result is a perturbative Hamiltonian
$H_{\rm pert} = H_1 + H_2 +\ldots$ written as a series with increasing
powers of $J$, and in which the impurity spin has disappeared. The
leading order $H_1$ is simply obtained by averaging the exchange Kondo
term over the spin up state,
\begin{equation}
  H_1 = \langle \uparrow \! | H_{\rm ex}  | \! \uparrow \rangle = 
\frac{J}{4} \sum_{\sigma,k,k'} \sigma c^\dagger_{k\sigma} c_{k'\sigma},
\end{equation}
corresponding to a spin-selective potential scattering term inducing
the phase shifts $\delta_{0\uparrow} = \pi - \pi \nu_0 J/4$ and
$\delta_{0\downarrow} = \pi \nu_0 J/4$. The next order, $H_2$,
arises from virtual impurity spin-flip process in which an
electron-hole pair (with opposite spins) is excited. It is obtained by
using the standard Schrieffer-Wolff technique, with the outcome
\begin{equation}
\label{eq:H2}
H_2  = -\frac{J^2}{8} \sum_{\{k_i\}} \frac{1}{B+\varepsilon_{3} - \varepsilon_{4}} c^\dagger_{k_1 \downarrow} c_{k_2 \uparrow} c^\dagger_{k_3 \uparrow} c_{k_4 \downarrow} + 
\text{h.c.}
\end{equation}
In order to compare this result with the FL form of
Eq.~\eqref{eq:H_FL}, we normal order the two equal-spin pairs of
  operators in \Eq{eq:H2} with respect to a reference ground state with
  spin-dependent chemical potentials
  $\ezerosigma = - \frac{1}{2}\sigma B_0$ close to $\muzerosigma = 0$
  (in the Kondo limit, where $\ed = - \infty$, there is no need to use
  $\ezero \neq 0$):
\begin{subequations}
\begin{align}
 c_{k_2 \uparrow} c^\dagger_{k_3 \uparrow} &
= - : \! c^\dagger_{k_3 \uparrow}  c_{k_2 \uparrow} \! : 
+\delta_{k_3,k_2} \theta(\varepsilon_{2}-\varepsilon_{0\uparrow}), \\
c^\dagger_{k_1 \downarrow} c_{k_4 \downarrow} &= 
:\! c^\dagger_{k_1 \downarrow} c_{k_4 \downarrow} \!:  
+ \delta_{k_1,k_4} \theta(\varepsilon_{0\downarrow}-\varepsilon_{1}) .
\end{align}
\end{subequations}
Inserting these expressions into Eq.~\eqref{eq:H2} yields, up
to a constant term, $H_2 = H_\alpha + H_\phi$, with
\begin{align}
\nonumber 
H_\alpha  = &  \frac{J^2 \nu_0}{8}\! \! \! \sum_{\sigma,k_1,k_2} \! \! \sigma   
\ln \! \left[ \frac{D}{B-B_0
+ \sigma (\varepsilon_1 \! - \! \varepsilon_{0\sigma}) }
\right]  c^\dagger_{k_1 \sigma} c_{k_2 \sigma}   \\
\label{eq:appendix-Halpha}
& \hspace{5.2cm} + \text{h.c.} \vspace{-5mm},
\\
  H_\phi   =  & \frac{J^2}{8} \! \sum_{\{k_i\}}
 \frac{:\!  c^\dagger_{k_3 \uparrow} c_{k_2 \uparrow} 
c^\dagger_{k_1 \downarrow}  c_{k_4 \downarrow} \! : }{B-B_0 +(\varepsilon_{3} - \ezero_\uparrow) -
(\varepsilon_{4}- \ezero_\downarrow)} 
+ \text{h.c.} , 
\end{align}
where $D$ is the high-energy cutoff of the Kondo model.  $H_\alpha$
describes elastic potential scattering. It can be expanded by assuming
$(\varepsilon_{1,2} - \ezerosigma) \ll B-B_0$. The zeroth order gives
the first logarithmic correction to the zero-energy phase shifts, 
\begin{align}
\label{eq:deltazerosigma-AppA}
\begin{Bmatrix}
\delta_{0\uparrow} \\ \delta_{0 \downarrow}
\end{Bmatrix}
= 
\begin{Bmatrix} \pi \\ 0 \end{Bmatrix}
\mp \frac{\pi \nu_0 J}{4} \mp  \frac{\pi (\nu_0
  J)^2}{4} \ln \left[\frac{D}{B-B_0}\right] . 
\end{align}
Changing from wavevector to energy summations, the first and second
orders obtained from \Eq{eq:appendix-Halpha}  reproduce
precisely~\footnote{up to terms which can be written as total
  derivatives in the action formalism, see Supplementary Note S-IV in
  Ref.~\onlinecite{mora2015}.} $H_\alpha$ in Eq.~\eqref{eq:H_FL}, with
$\bar{\alpha}_1 = 0$, $\alpha_2 = 0$ and
\begin{equation}\label{eq:alpha}
\frac{\alpha_1}{\pi} = \frac{(\nu_0 J)^2}{4(B- B_0)}, 
\qquad \frac{\balpha_2}{\pi} = - \frac{(\nu_0 J)^2}{8 (B-B_0)^2}.
\end{equation}
Next, expand $H_\phi$ to first order in
$(\varepsilon_{3,4} - \ezerosigma)/(B-B_0)$. The result coincides with
$H_\phi$ in Eq.~\eqref{eq:H_FL}, with $\phi_2 = 0$,
\begin{equation}\label{eq:phi}
\frac{\phi_1}{\pi} = \frac{(\nu_0 J)^2}{4 (B-B_0)}, 
\qquad \frac{\bphi_2}{\pi} = - \frac{(\nu_0 J)^2}{2 (B- B_0)^2}.
\end{equation}
Eqs.~\eqref{eq:deltazerosigma-AppA} 
  to \eqref{eq:phi} are the main results of this appendix. They
  explicitly give all the FL parameters in terms of the bare
  parameters of the Kondo model and the dummy reference energies
  $\ezerosigma = -\frac{1}{2} \sigma B_0$, illustrating explicitly
  that the latter occur only in the combination $B-B_0$ [cf.\
  Eq.~\eqref{eq:invariance-with-shifts}]. It is easy to verify
  explicitly that Eqs.~\eqref{eq:deltazerosigma-AppA} 
  to \eqref{eq:phi} satisfy the FL relations
  \eqref{eq:FL-relations-I} (in the latter, all derivatives w.r.t.\
  $\ed$ vanish in the Kondo limit). Moreover, the above derivation
  clarifies the underlying reason for why the FL parameters
  necessarily must be mutually interrelated: they arise as expansion
  coefficients of the actual physical Hamiltonian in the
  large-field Kondo limit, namely $H_2$ 
  of Eq.~\eqref{eq:H2}, whose functional form fully fixes
  all terms in the expansion $H_2 = H_\alpha + H_\phi + \dots$.

  Eqs.~\eqref{eq:deltazerosigma-AppA} 
  to \eqref{eq:phi} can also be used to test our predictions
  \eqref{relations-kondo} for how the FL parameters  are related to
  susceptibilities. To this end, we remove the dependence on the dummy
  reference energy by setting it to $\ezerosigma = \muzerosigma = 0$
  [as done in Eq.~\eqref{friedel2}].  Then, we directly compute the 
magnetization and spin susceptibility at large magnetic field. The
Bethe Ansatz solution provides a universal expression for the
magnetization of the Kondo model,
\begin{equation}\label{magnet-kondo}
  m_d = \frac{1}{2} - \frac{1}{2 \pi^{3/2}}
  \int_0^{\infty} d t  \, \frac{\sin( \pi t) (\beta_r t)^{-t}}{t}  \, \Gamma \left( \tfrac{1}{2} +t \right) 
\end{equation}
where the ratio
$\beta_r = \frac{\pi }{8} (B/\Tkchi)^2 \ge \frac{1}{e} $
is written in terms of the Kondo temperature $\Tkchi$ extracted from
the zero-field spin susceptibility [\Eq{eq:tksus}]. 
At large magnetic
fields, Eq.~\eqref{magnet-kondo} can be expanded in powers of
$1/\ln \beta_r$.  Noting that (we use that
$\Tkchi \sim D \, e^{-1/\nu_0 J}$)
\begin{equation}
\frac{2}{\ln \beta_r} \simeq \frac{\nu_0 J}{1 + \nu_0 J \ln (B/D)} \simeq \nu_0J + \ldots
\end{equation}
we also find an expansion for $\nu_0 J \ll 1$. At large magnetic
fields, one obtains
\begin{equation}\label{eq:largeb}
m_d = \frac{1}{2} - \frac{\nu_0 J}{4}, \quad \chi_s = \frac{(\nu_0 J)^2}{4 B}, \quad \frac{\partial \chi_s}{\partial B} = -\frac{(\nu_0 J)^2}{4 B^2}.
\end{equation}
Inserting these susceptibilities into Eqs.~\eqref{relations-kondo}
for the FL parameters, we recover Eqs.~\eqref{eq:alpha},~\eqref{eq:phi},
which serves as a nice consistency 
check for Eqs.~\eqref{relations-kondo}.
[\Eq{eq:largeb}
for $\md$ does not strictly approach $\frac{1}{2}$ in the limit
$B \to \infty$, because the calculation is perturbative in $\nu_0 J$.]

In summary, in this appendix we explicitly derived the FL Hamiltonian
at large magnetic field and checked the FL relations advertised in this
paper.


\section{RPT-FL dictionary}
\label{app:RPT}

It is instructive to relate the FL parameters introduced in this work
to the parameters that are used in renormalized perturbation theory
(RPT)
\cite{hewson1997kondo,hewson2005b,hewson1993,*hewson1993b,*hewson1994,*hewson2001,*hewson2004,*hewson2006b,*bauer2007,*edwards2011,*edwards2013,*Pandis2015,hewson2005,hewson2006}
to parametrize the low-energy behavior of the retarded local Green's
function of the impurity,
$G_{d\sigma} (\omega) = 1/[\omega - \Sigma_\sigma(\omega)]$.  If
$\omega$, $T$ and $eV$ are so small that the impurity self-energy may be
expanded to second order in these variables, this correlator can be
expressed in the form
\label{subeq:G-general}
\begin{align}
\label{eq:G-general}
  G_{d\sigma} (\omega) & = \frac{\tzs}{\omega - \tes + i 
  \tds + \tRs + i \tIs }.
\end{align}
All parameters carrying tildes are understood to be functions of
magnetic field.
$\tzs = \left[1- \partial_\omega \Sigma'_\sigma (0)\right]^{-1}$ is
the quasiparticle weight,
$\tes = \left[\varepsilon_{d \sigma} + \Sigma_{\sigma}(0)\right] \tzs$
the renormalized position of the local level with spin $\sigma$, and
$\tds = \tzs \Delta$ its renormalized width.  $\tRs$ and $\tIs$ are
the real and imaginary parts of $-\Sigma_\sigma^{(2)} \tzs$, coming from the
second-order term in the self-energy, which we parametrize as
\begin{subequations}
\begin{align}
\label{eq:tRs}
\tRs & = \tRso \omega^2 + 
\tRsV \bigl[\tfrac{1}{3} (\pi T)^2 + \tfrac{1}{4} (eV)^2\bigr] , 
\\
\label{eq:tRs2}
\tIs & = \tIso \bigl[\tfrac{1}{3} \omega^2 + 
\tfrac{1}{3} (\pi T)^2 + \tfrac{1}{4} (eV)^2\bigr]  ,
\end{align}
\end{subequations}
where $\tRso$, $\tRsV$ and $\tIso$ are constants independent of
$\omega$, $T$ and $eV$.  The imaginary part of the second-order
self-energy can only depend on the combination of energy, temperature
and bias stated in Eq.~\eqref{eq:tRs2}, because it is governed by the
second-order term of the inelastic $T$-matrix, which we know to depend
only on this combination [Eq.~\eqref{eq:T-inelastic}].  The
corresponding real part, however, requires two separate coefficients for
its energy dependence and its temperature and voltage
dependence [Eq.~\eqref{eq:tRs}], because the former also receives a
contribution from the elastic $T$-matrix, but the latter does not.

The spin-resolved version of the FL spectral function discussed in the
main text is normalized such that $A_\sigma(0) = 1$ for the symmetric
Anderson model at $T=V=B=0$.  It is related to the imaginary part of
the local Green's function by
$A_\sigma(\omega) = - (\pi \Delta) \frac{1}{\pi} \text{Im}
G_\sigma(\omega)$, hence [from Eq.~\eqref{eq:G-general}]:
\begin{align}\label{spec-jvd}
A_\sigma (\omega)   & =   \dfrac{\tds(\tds + \tIs)}{
( \omega - \tes  + \tRs)^2 + (\tds + \tIs)^2}  \, . 
\end{align}
When this expression is expanded in the form of the spin-resolved
versions of Eq.~\eqref{eq:expandAepsilon-general}, 
 \begin{align}  
A_\sigma (\omega)   =  
A_{0\sigma}  +  A_{1\sigma} \omega 
\label{eq:expandAepsilon-general-app}
    - \tCAs \! \left[\tfrac{1}{3} (\pi T)^2 \! + \!  
 \tfrac{1}{4}(e V)^2\right] -  \CAs \omega^2 , & 
\end{align}
and the expansion coefficients are expressed in terms
of 
\begin{align}
\trhos & = -\frac{1}{\pi} \text{Im}G_{d\sigma}(0)  = 
\frac{\tds/\pi}{\tds^2 + \tes^2} \, ,
\\
\sin( \delta_{0\sigma}) & = \frac{\tds}{\sqrt{\tds^2+ \tes^2}}   , 
\;  \; 
\cos (\delta_{0\sigma})  = \frac{\tes}{\sqrt{\tds^2+ \tes^2}}  , 
\end{align}
one readily obtains: 
\begin{subequations}
\label{compare-coefficients}
\begin{align}
\label{eq:A0-identify}
A_{0 \sigma}  = {} &  \sin^2  (\delta_{0\sigma}) \, , 
\\
\label{eq:A1-identify}
 A_{1\sigma}  =  {} &   \pi \trhos \sin(2 \dels) \, , 
\\
\label{eq:tildeCA-identify}
\tCAs  = {} & 
- \pi \trhos \bigl[\tRsV \sin(2 \dels) + \tIso \cos(2 \dels) \bigr] \, , 
\\ 
\nonumber
\CAs  = {} &  - \pi \trhos \bigl[\tRso \sin(2 \dels) + 
\tfrac{1}{3} \tIso \cos(2 \dels) \bigr] 
\\ &  +  (\pi \trhos)^2 [4 \sin^2 (\dels)  - 3 ] \, . 
\label{eq:CA-identify}
\end{align}
\end{subequations}
By comparing Eqs.~\eqref{compare-coefficients} to  the expressions 
\eqref{subeq:Acoefficients}  of the
main text, we can express all the FL parameters in terms
of RPT parameters. Eqs.~\eqref{eq:A1-identify} and \eqref{eq:tildeCA-identify}
imply
\begin{subequations}
\label{eq:FL-RPT-general-dictionary}
\begin{align}
\label{eq:FL-RPT-general-dictionary-1}
\alpha_{1\sigma} & = \pi \trhos \, , 
\quad 
\phi_1  = \sqrt{\tfrac{2}{3} \pi \trhos \tIso} \, , 
\quad  
 \phi_{2 \bar \sigma}  = - 4 \pi \trhos \tRsV \, .
\end{align}
Inserting these into  
Eq.~\eqref{eq:CA-main-paper} and solving for $\alpha_{2\sigma}$ we find:
\begin{align}
\label{eq:FL-RPT-general-dictionary-2}
 \alpha_{2\sigma} & = 
(\pi \trhos)^2 \cot (\dels) + 
\pi \trhos \tRso \, . 
\end{align}
\end{subequations}
Eqs.~\eqref{eq:FL-RPT-general-dictionary} constitute a useful 
dictionary that relates the RPT parameters, which characterize
the impurity dynamics, to the FL parameters, which characterize
the quasiparticle dynamics.   

In conjunction with Eqs.~\eqref{fermic}, the RPT-FL dictionary can be
used to express the RPT parameters in terms of local ground state
susceptibilities; they can thus be computed exactly via the Bethe
Ansatz.  Moreover, if alternative strategies (e.g.\ NRG) are used to
compute the RPT parameters, then the relations \eqref{eq:odd-zeroB} to
\eqref{eq:large-B} between various FL parameters that hold  for certain special
cases (zero field, or particle-hole symmetry, or the Kondo limit),
suitably transcribed using the RPT-FL dictionary, provide useful
consistency checks on the RPT parameters.

\section{Agreement of the transport coefficients $C_T$ and $C_V$ 
 with those of Oguri and Hewson}\label{app:oh}

In this Appendix we verify that our expressions for
the transport coefficients $C_T$ an $C_V$, given by
Eqs.~\eqref{eq:cT-cV-cA-ctildeA} and \eqref{subeq:Acoefficients},
are consistent with those recently derived by Oguri and Hewson
\cite{Oguri2017,*Oguri2017a,*Oguri2017b}. Their
definition of the transport
coefficients occuring in our conductance expansion~\eqref{eq:c's-B-dependent}
reads
\begin{equation}
G(V,T)=\frac{2e^2}h\sum_\sigma\Big[\sin^2(\delta_{0\sigma})-\bar C_{T,\sigma} T^2-\bar C_{V,\sigma}(eV)^2\Big]\,,
\end{equation}
with
\begin{equation}\label{eq:oh}
\begin{split}
\bar C_{T,\sigma}=\frac{\pi^4}{6}&\Bigg[-\cos(2\delta_{0\sigma})\big(\chi^2_{\sigma\sigma}+2\chi^2_{\uparrow\downarrow}\big)\\&+\frac{\sin(2\delta_{0\sigma})}{2\pi}\left(\frac{\partial\chi_{\sigma\sigma}}{\partial \varepsilon_d}+\sigma\frac{\partial \chi_{\uparrow\downarrow}}{\partial h}\right)\Bigg]\,\\
\bar C_{V,\sigma}=\frac{\pi^2}{8}&\Bigg[-\cos(2\delta_{0\sigma})\big(\chi^2_{\sigma\sigma}+5\chi^2_{\uparrow\downarrow}\big)\\&+\frac{\sin(2\delta_{0\sigma})}{2\pi}\left(\frac{\partial\chi_{\sigma\sigma}}{\partial \varepsilon_d}+\frac{\partial\chi_{\uparrow\downarrow}}{\partial \varepsilon_d}+2\sigma\frac{\partial \chi_{\uparrow\downarrow}}{\partial h}\right)\Bigg]\,,
\end{split}
\end{equation}
in which $2h=B$ and $\chi_{\sigma\sigma'}=-\partial n_{d\sigma}/\partial\varepsilon_{d\sigma'}$. Notice that $\chi_{\sigma\sigma'}=\chi_{\sigma'\sigma}$, as $n_{d,\sigma}(\varepsilon_d,h)=n_{d,-\sigma}(\varepsilon_d,-h)$. In this notation, after making the substitutions 
\begin{align}\label{eq:derivatives}
\frac{\partial}{\partial\varepsilon_d}&=\frac{\partial}{\partial\varepsilon_{d\uparrow}}+\frac{\partial}{\partial\varepsilon_{d\downarrow}}\,, &\frac{\partial}{\partial h}&=-\frac{\partial}{\partial\varepsilon_{d\uparrow}}+\frac{\partial}{\partial\varepsilon_{d\downarrow}}\,, 
\end{align}
our susceptibilities~\eqref{eq:susceptibilities-largeB} read  
\begin{equation}\label{eq:chioh}
\begin{split}
\chi_m=\frac{\chi_{\uparrow\uparrow}-\chi_{\downarrow\downarrow}}2&\,,\qquad
\chi_s=\frac{\chi_{\uparrow\uparrow}+\chi_{\downarrow\downarrow}-2\chi_{\uparrow\downarrow}}4\,,\\
\chi_c&=\chi_{\uparrow\uparrow}+\chi_{\downarrow\downarrow}+2\chi_{\uparrow\downarrow}\,.\
\end{split}
\end{equation}
Comparing Eqs.~\eqref{eq:oh} with our
  Eqs.~\eqref{subeq:Acoefficients} and \eqref{eq:cT-cV-cA-ctildeA}
  for $C_T$ and $C_V$, one finds that they are consistent  provided
  that the following relations hold:
\begin{align}
\label{eq:I}\pi^2(\chi^2_{\sigma\sigma}+2\chi^2_{\uparrow\downarrow})&=\alpha_{1\sigma}^2+2\phi_1^2\,,\\
\label{eq:II}\pi^2(\chi^2_{\sigma\sigma}+5\chi^2_{\uparrow\downarrow})&=\alpha_{1\sigma}^2+5\phi_1^2\,,\\
\label{eq:III}\frac\pi2\left[\frac{\partial\chi_{\sigma\sigma}}{\partial \varepsilon_d}+\sigma\frac{\partial \chi_{\uparrow\downarrow}}{\partial h}\right]&=\frac14\phi_{2\bar\sigma}-\alpha_{2\sigma}\,, \\
\label{eq:IV}\frac\pi2\left[\frac{\partial\chi_{\sigma\sigma}}{\partial \varepsilon_d}+\frac{\partial\chi_{\uparrow\downarrow}}{\partial \varepsilon_d}+2\sigma\frac{\partial \chi_{\uparrow\downarrow}}{\partial h}\right]&=\frac34\phi_{2\bar\sigma}-\alpha_{2\sigma}\,. 
\end{align}
This is indeed the case.
Equalities~\eqref{eq:I} and~\eqref{eq:II} readily follow from the substitution 
\begin{align}\label{eq:translation}
\phi_1&=-\pi\chi_{\uparrow\downarrow}\,,&\alpha_{1\sigma}=\sigma\pi\chi_{\sigma\sigma}\,.
\end{align}
Equation~\eqref{eq:III} is shown by writing its right hand side in the
form
$\phi_{2\bar\sigma}/4-\alpha_{2\sigma}=\pi(\partial_{\varepsilon_{d\sigma}}\chi_{\sigma\sigma}+\partial_{\varepsilon_{d,-\sigma}}\chi_{\uparrow\downarrow})/2$,
then by applying the substitution~\eqref{eq:derivatives} on the left
hand side of Eq.~\eqref{eq:III}, and using 
$\partial_{\varepsilon_{d\sigma}}\chi_{-\sigma,-\sigma}=\partial_{\varepsilon_{d,-\sigma}}\chi_{\sigma,-\sigma}$. The
equality~\eqref{eq:IV} is shown in a similar fashion: applying
Eq.~\eqref{eq:chioh} on the right hand side, the equality is derived
provided that
$\partial_{\varepsilon_d}\chi_{\sigma\sigma}+\sigma\partial_h\chi_{\uparrow\downarrow}-(\partial_{\varepsilon_d}-\sigma\partial_h)(\chi_{\uparrow\uparrow}+\chi_{\downarrow\downarrow})/2=0$,
which is readily shown after substitution~\eqref{eq:derivatives}.

\section{NRG computation of $\CT$ and $\CV$} \label{app:NRG}

In this appendix we describe a consistency check for our FL theory. We
consider the case of particle-hole symmetry ($\varepsilon_d = - U/2$),
and use the numerical renormalization group (NRG)
\cite{Wilson1975,Bulla2008} to compute the equilibrium impurity
spectral function, $A(\epsilon)$, as function of temperature and
magnetic field. Although this is an equilibrium quantity, it contains
sufficient information to determine not only $C_T$ but also $C_V$ ---
the point is that both these transport coefficients are fully
determined by the expansion coefficients $\tCA$ and $\CA$ [see
\Eq{eq:expandAepsilon-general}], which can be extracted from the
low-energy behavior of the equilibrium spectral function
$A(\epsilon)$.  As shown below, the NRG results for the magnetic-field
dependence $C_T$ and $C_V$ are consistent with the FL predictions
obtained in the main text.

\subsection{Method, definitions and conventions} 

We employ the full-density-matrix NRG
approach (fdm-NRG) \cite{Weichselbaum2007,Weichselbaum2012a},  based on
complete basis sets \cite{Anders2005}.
We also fully exploit non-abelian symmetries \cite{Weichselbaum2012b}
where applicable. Here this is SU(2) spin when there
is no magnetic field ($B=0$), and SU(2) particle-hole
symmetry. During NRG iterative diagonalization we keep track
of symmetry multiplets rather than individual states, and
\Nkept specifies the reduced, i.e.\ effective dimensionality
in terms of number of kept multiplets.
%
The discrete spectral data is $z$-averaged \cite{Oliveira92,Zitko09}
by averaging over $n_z$ equally spaced $z$-shifts, with $z \in (0, 1]$,
for the logarithmically discretized
conduction band energies, $\Lambda^{-n-z}$, 
with the discretization parameter $\Lambda\gtrsim2$
and $n$ being integers.

At particle-hole symmetry and in equilibrium,
  \Eq{eq:expandAepsilon-general} for the  low-energy behavior of the
  impurity spectral function simplifies to
\begin{align}
\label{eq:expandAepsilon-appendix}
& A (\varepsilon) \equiv \sum_\sigma A_\sigma(\epsilon) 
=   A_0  - \tfrac{1}{3} \tCA  (\pi T)^2  -  \CA \ve^2 .
\end{align}
The output of an NRG computation of $A(\ve)$ yields 
a representation of this function as 
a weighted sum over discrete delta functions.
Usually these are broadened individually to obtain 
a smooth, continuous curve \cite{Bulla2008,Weichselbaum2007,Lee2016}.
The coefficients in \Eq{eq:expandAepsilon-appendix}
may then be obtained by fits to such broadened curves.
However, a much better way to extract these coefficients
is from integrals of the discrete data itself --
this avoids the need for broadening and hence
yields significantly lower error bars on the Fermi
liquid coefficients \cite{Weichselbaum2007,Hanl2014}.
A convenient way of extracting $\tCA$ and $\CA$ from
the discrete NRG 
data is to consider the integral 
\begin{align}
  g_\alpha(T) & \equiv
  \tfrac{1}{2} \int d\epsilon A(\epsilon) \frac{-df_{\alpha}(\epsilon)}{d\epsilon} \, , \label{eq:MWcond}
\end{align}
where $\tfrac{1}{2} A$ is the spin-averaged spectral function, and
$f_\alpha(\omega) = [1+e^{\omega/(\alpha T)}]^{-1}$ 
a Fermi function with temperature scaled by a factor  $\alpha$. 
For $\alpha = 1$, \Eq{eq:MWcond} 
corresponds to 
the dimensionless version of the linear
conductance of \Eq{eq:c's-B-dependent},
$g(T) \equiv g_1(T) = G(0,T)/G_0$. 
Inserting \Eq{eq:expandAepsilon-appendix} into \Eq{eq:MWcond} we obtain 
\begin{eqnarray}
g_\alpha (T) \equiv 
\tfrac{1}{2} A_0 - C_\alpha T^2 , \ \text{with }  
C_\alpha  = \tfrac{\pi^2}{6} (\tCA + \alpha^2 \CA) \, . \qquad
\end{eqnarray}
The coefficient $C_\alpha$ can be extracted from the curvature
of $g_\alpha(T)$ in the limit $T/\Tk \ll 1$, 
and $C_T$ or $C_V$ are found, via \Eq{eq:cT-cV-cA-ctildeA}, 
by doing this for $\alpha = 1$ or $\alpha = 1/\sqrt{3}$:
\begin{subequations}
\label{eq:cT-cV-cA-ctildeA-appendix} 
\begin{align}
\CT  & = \tfrac{1}{6}\pi^2 (\tCA + \CA) = C_1\, , \\
\CV  & = \tfrac{3}{8}(\tCA + \tfrac{1}{3}\CA ) = 
\bigl(\tfrac{3}{2\pi}\bigr)^2 C_{1/\sqrt{3}} \, . 
\end{align}
\end{subequations}
In the Kondo limit $U/\Delta \to \infty$, the zero-field
values of these coefficients follow from \Eq{subeq:CA-tCA-Kondo-limit},
\begin{align}
C_1^K 
= C_T^K = \frac{\pi^4}{16 \Tk^2} , 
\qquad C_V^K =  \frac{3 \pi^2}{32 \Tk^2} \, .  
\label{eq:Calpha:K}
\end{align}
with $\Tk \equiv 1/4\chi^{B=0}_s$
defined from the zero-field impurity spin susceptibility.
\begin{figure}[t]
  \includegraphics[width=1\linewidth]{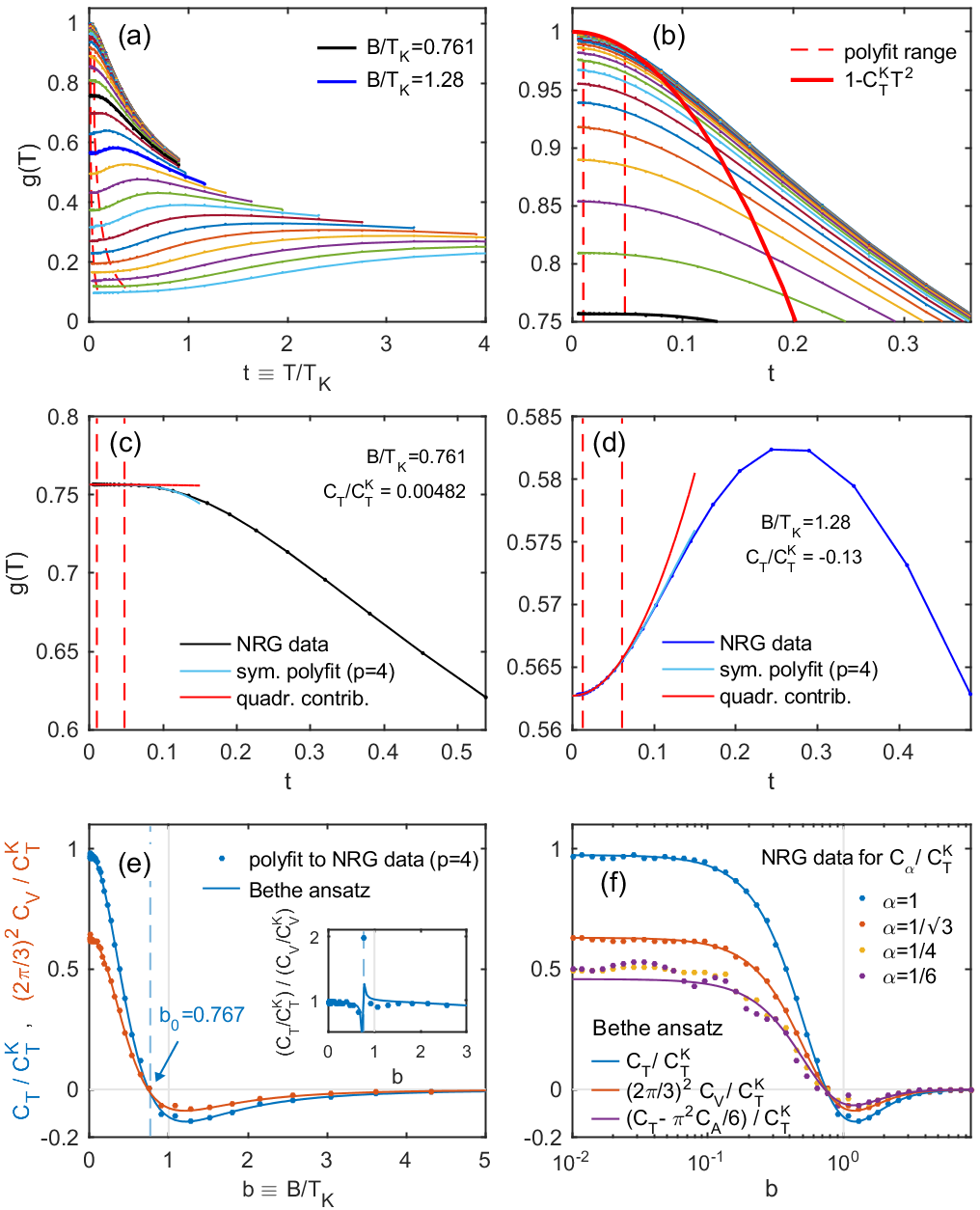}
  \caption{
NRG analysis 
   of the transport coefficients 
     $\CT$ and $\CV$. 
 (a) The linear
    conductance, $g(T) = g_{\alpha = 1}(T)$, plotted 
as function of $t \equiv T/\Tk$, for a uniform
logarithmic grid of magnetic fields around $\Tk$
[cf.\ data points in panels (e-f)].
The red vertical
    dashed lines indicate the fitting range used to 
    extract curvatures.
    (b) Zoom into the low-temperature regime of (a).  The thick red
    line shows the FL prediction at zero field, $1-C_T^K
    T^2$. 
    (c,d) Illustration of the fit quality for two curves from (a) for
    magnetic fields around $C_T\approx 0$ and around the largest
    negative value of $C_T$, respectively.  
(e) $C_1 = \CT$ 
(blue)
    and  $C_{1/\sqrt 3} = (2 \pi/3)^2 \CV$ (red), normalized
by $\CT^K$ and plotted versus
    $B/\Tk$. The agreement with the corresponding FL predictions
    (solid lines in matching color) 
    from the main text is excellent. Inset: the ratio
 $\CT/\CT^K$ over $\CV/\CV^K$, showing that both 
have the same limits for $B/\Tk \ll $ and $\gg 1$.
    (f) $C_\alpha/\CT^K$ for several values of $\alpha$, 
plotted on a log scale as function of $B/\Tk$.
The data for $\alpha = 1$ and $1/\sqrt{3}$
replicate those from from (e) for $C_T$ and $C_V$.
For comparison, solid lines show Bethe ansatz data.
      NRG
    parameters: $\Lambda=2$, $z$-averaged at $n_z=8$; truncation by
    number of multipets, with 
    $\Nkept=1060$ (corresponding to $6944$ states).}
\label{fig:CTU5}
\end{figure}


\subsection{Numerical results}

Our NRG calculations were performed for the
single-impurity Anderson model with
a box-shaped density of states with half-bandwidth
$D=1$, which thus sets the unit of energy unless
specified otherwise.
For the results shown in Fig.~\ref{fig:CTU5}, we
used a hybridization strength $\Delta=10^{-3}$, 
$U/\Delta=5$, $\epsilon_d=-U/2$,
resulting in a Kondo temperature of $\Tk=2.990\cdot 10^{-4}$.
Since both $U$ and $\Gamma$ are much smaller
than the bandwidth, and
the local dynamics is cut off by the energy scale $U$,
the parameter regime analyzed essentially is
equivalent \cite{Hanl2014} to an infinite bandwidth scenario
as considered in the main text.

Our NRG results for the temperature-dependent linear conductance, 
$g(T) = g_{\alpha=1}(T)$,  are shown in Fig.~\ref{fig:CTU5}(a), 
for a fine-grained set of 
$B$-values 
uniformly distributed around $\Tk$ on a logarithmic grid.
To determine
the Fermi liquid coefficient $C_T = C_{\alpha =1}$ for a given 
field, the curvature of the corresponding curve 
must be extracted in the regime $t \equiv T/\Tk \ll 1$. On the other
hand, $t$ must be sufficiently large
that the temperature-dependent variation in $g(T)$
exceeds the NRG resolution (typically, $\delta g
\gtrsim 10^{-3}$). In practice, we constrain the
fitting interval to $t \in [0.01,0.05] \times \max(\Tk,B)$
[vertical dashed lines in Figs.~\ref{fig:CTU5}(b-d)].
Since we use a fine-grained logarithmic grid for $t$,
cutting out $t<0.01$ is also required to avoid
a bias of the fit towards a dense set of data
points at $t=0$.
We perform a symmetrized $(p=4)$-th order polynomial 
fit, by also including the mirror image $t \to -t$
of the data. This way, only even polynomial coefficients
are non-zero.

Two exemplary fits are shown in Figs.~\ref{fig:CTU5}(c,d).
In panel (c), $C_T\approx 0$, 
indicating that the field is close to 
the value $B_T$ where the curvature $C_T$ changes sign.
In contrast, Fig.~\ref{fig:CTU5}(d) analyzes the
conductance data for the magnetic field where the
upward curvature in the conductance is maximal
[this corresponds to the minimum of $c_T$ in
Fig.~\ref{fig:CTU5}(e)]. Both panels (c)
and (d) show that $g(T)$ deviates significantly from
the quadratic regime already for $t \simeq 0.1$. Hence
smaller $t$ is required for the fit (here we use
$t\leq 0.05$) while already also allowing
for quartic corrections. 
Moreover, note that $g_\alpha(T)$ varies by $\lesssim 0.5\%$ 
over the fitting interval, illustrating that an accurate
determination of the curvature requires $g_\alpha(T)$ to be known with
an accuracy of order $\lesssim 0.1\%$. 

Fig.~\ref{fig:CTU5}(e) shows the curvatures $\CT$
and $(2 \pi/3)^2 \CV$ extracted from all the curves in panel (a),
normalized by $\CT^K$ and plotted as dots, 
as functions of $B/\Tk$. The NRG results are in very good agreement
with the FL predictions (Bethe ansatz data, solid lines,
replotted from Fig.~3 of main
text). 
For the specific value $U/\Delta=5$ used in this plot,
the field at which $C_T$ vanishes, as
obtained by interpolation from the discrete set of NRG data points, is
$B_T /\Tk =0.767$.
This agrees to within about 1\% with the FL
prediction.
Panel (f) replots the data from (e)
on a log-$x$ scale, showing good agreement down to the smallest
magnetic fields.

\begin{figure}
  \includegraphics[width=0.96\linewidth]{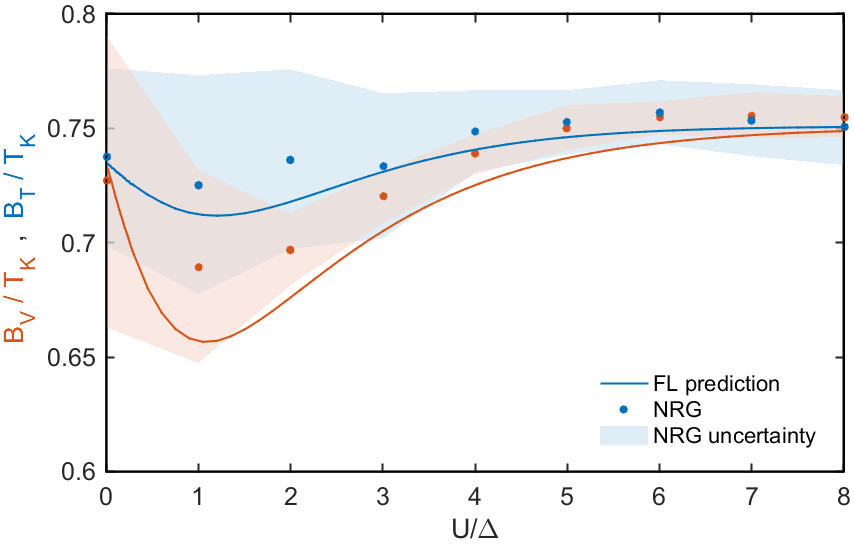}
  \caption{Comparison of results from NRG and FL theory for $B_T$
      and $B_V$, corresponding to Fig.~\ref{fig-vanishing} of the main
      text.  The color-matched shaded regions
      are estimates
      for the error margins, simply based on matlab polyfit confidence
      intervals. NRG parameters: $\Delta=0.001$, $\Lambda=2$, with
      $\Nkept\leq 1244$ multiplets ($8176$ states), $z$-averaged at
      $n_z=4$.}\label{fig:B0Fig3}
\end{figure}

We have repeated the above analysis for a range of interaction
strengths to extract $B_T$ and $B_V$ as functions of $U/\Delta$. 
The results are shown in Fig.~\ref{fig:B0Fig3}. The NRG results
are consistent with the FL predictions within the indicated
simple error estimates. Note that  the error estimates 
increase with decreasing $U/\Delta$, because this 
causes the Kondo peak to become broader, making it
more difficult to accurately extract curvature coefficients.
The fact that the error estimates are quite sizeable,
varying between 3\% and 12\%, reflects the challenge
of extracting curvatures from an energy window
that necessarily has to be very small to satisfy
the requirement  $\ve,T \ll \Tk$.

We conclude this subsection with a technical remark.
Since $\tCA = \frac{6}{\pi^2} C_0$ and
$\CA = \frac{6}{\pi^2} (C_1 - C_0)$, one may attempt 
to compute these coefficients via $C_1$ and $C_0$. 
Note, though, that $C_0$ can not be
extracted by simply using $\alpha \to 0^+$,  because the discrete
NRG data looses spectral support
at energies below $\sim T/10$
(see Fig.~2(b) of \cite{Weichselbaum2007}).  Therefore, when 
the peaked function $-d f_\alpha(\ve)/d\ve$ becomes too narrow
as $\alpha$ is decreased, the
numerically determined $g_\alpha(T)$ first becomes a noisy function of $T$,
and eventually drops to zero for very small $\alpha$. Indeed,
Fig.~\ref{fig:CTU5}(f), which displays curves of $C_\alpha/\CT^K$
versus $B/\Tk$ for several values $\alpha \in [0,1]$, shows that these
curves become more `noisy' as $\alpha$ is decreased.  The increase in
noise in spectral resolution seen at $\alpha=1/4$ becomes large very
quickly for $\alpha<0.25$.

$C_0$ can  nevertheless be determined  to fairly good
accuracy  by noting that 
the overall shape of the curves in Fig.~\ref{fig:CTU5}(f)
essentially stops changing for $\alpha < 0.25$. 
Hence $C_{0.25}$ may be viewed as an approximation for $C_0$.
Indeed, its low-field limit, $C_{0.25} \simeq C_T^K /2$,
is consistent with the value expected for 
$C^K_0  = C_T^K/2$. 
For comparison, Fig.~\ref{fig:CTU5}(f) also shows the Bethe
ansatz data for 
$C_0 
= 
C_1 - \tfrac{\pi^2}{6} \CA 
= 
C_T - \tfrac{\pi^2}{6} \CA .
$
The agreement with the NRG data is within the increased
level of spectral noise of the NRG data.
 
\subsection{Remarks on our previous NRG results}
\label{app:previousNRG}

We conclude this appendix with a brief discussion of why, in
retrospect, the NRG results presented in a previous version of this
paper, \cite{Filippone2017}, were unreliable. There are two main
reasons: first, they were obtained using only 1024 states (here we use
6944), and no $z$-averaging (here we use $n_z = 8$). Hence they did
not achieve the $\mathcal{O}(0.1\%)$ accuracy needed for
$g_\alpha (T)$ to allow an accurate determination of its curvature.
Second, we had attempted to extract the curvature coefficients 
$\CA$ and $\tCA$ by following a strategy described in
section~IV.D of Ref.~\cite{Hanl2014}. However, that
strategy had been devised only to determine $\CA$ at $T{=}0$.
In retrospect, our attempt to generalize it to determine both
$\CA$ and $\tCA$ at nonzero $T$ had been flawed, which is
why we now instead use $C_1$ and $C_{1/\sqrt{3}}$ to determine
$C_T$ and $C_V$, as described above.

\section{Perturbation results in the  $\varepsilon_d\gg\Delta$ limit.}
\label{app:largeed}
The FL coefficients needed to derive the spectral coefficients
$\tcA,\cA$ and transport coefficients $c_V, c_T$, depend on the
zero-temperature dot occupation functions $\ndsigma$ and their
derivatives with respect to the level energy $\ed$ and magnetic field
$B$.  Deep in the empty-orbital regime, where $\varepsilon_d\gg\Delta$, 
the leading corrections to the noninteracting 
occupations, 
\begin{align}
n^0_{d \uparrow} =& \tfrac{ 1}{ 2} - \tfrac{1}{\pi} \, {\rm arctan} 
\left[\left(\varepsilon_d-\tfrac{1}{2}B \right)/\Delta\right] , 
\end{align}
can be computed perturbatively in
the dot-lead hybridization, using the spin-down state of the dot as
intermediate state (see also Ref.~\onlinecite{filippone2013}), with the
result 
\begin{subequations}
\begin{align}
n_{d \uparrow}=& n^0_{d \uparrow} -\frac\Delta\pi \frac{U
n_{d \uparrow}^0 (1- n_{d \uparrow}^0)}{(U+\varepsilon_d+\frac{1}{2}B)(\varepsilon_d+ \frac{1}{2}B )},\\
n_{d \downarrow} =&\frac\Delta\pi\left(\frac{1-
n_{d \uparrow}^0 }{\varepsilon_d+\frac{1}{2}B}+
\frac{n_{d \uparrow}^0}{\varepsilon_d+U+\frac{1}{2}B}\right) . 
\end{align}
\end{subequations}
All FL parameters can be straightforwardly computed from these
expressions using \Eqs{eq:susceptibilities-largeB} and \eqref{fermic}.
To compare with the numerical results in Fig.~\ref{fig:cvcuts}, we
substitute $\varepsilon_d \to \ted$ [cf. Eq.~\eqref{eq:edtilde}]
in the final result for $c_V$.

\bibliographystyle{apsrev4-1}
\bibliography{biblio}
 
\end{document}